\documentclass[journal=jctcce,manuscript=article]{achemso}

\usepackage[version=3]{mhchem} 
\usepackage{amsmath}
\usepackage{amssymb}
\usepackage{braket}
\usepackage{xcolor} 
\usepackage{pgf}
\usepackage{chemformula}
\usepackage{subfig}
\usepackage{graphicx}
\usepackage[final]{changes}
\usepackage{cancel}
\usepackage{hyperref}
\author{Håkon Emil Kristiansen}
\email{h.e.kristiansen@kjemi.uio.no}
\author{Benedicte Sverdrup Ofstad}
\author{Eirill Hauge}
\affiliation[Hylleraas Centre]
{Hylleraas Centre for Quantum Molecular Sciences, Department of Chemistry, University of Oslo, N-0315 Oslo, Norway}
\alsoaffiliation[Simula]{Simula Research Laboratory, Kristian Augusts gate 23, 0164 Oslo, Norway}
\author{Einar Aurbakken}
\affiliation[Hylleraas Centre]
{Hylleraas Centre for Quantum Molecular Sciences, Department of Chemistry, University of Oslo, N-0315 Oslo, Norway}
\author{Øyvind Sigmundson Sch{\o}yen}
\affiliation{Department of Physics, University of Oslo, N-0316 Oslo, Norway}
\author{Simen Kvaal}
\affiliation[CAS]
{Centre for Advanced Study at the Norwegian Academy of Science and Letters, Drammensveien 78, N-0271 Oslo, Norway}
\alsoaffiliation[Hylleraas Centre]
{Hylleraas Centre for Quantum Molecular Sciences, Department of Chemistry, University of Oslo, N-0315 Oslo, Norway}
\author{Thomas Bondo Pedersen}
\affiliation[CAS]
{Centre for Advanced Study at the Norwegian Academy of Science and Letters, Drammensveien 78, N-0271 Oslo, Norway}
\alsoaffiliation[Hylleraas Centre]
{Hylleraas Centre for Quantum Molecular Sciences, Department of Chemistry, University of Oslo, N-0315 Oslo, Norway}
\email{t.b.pedersen@kjemi.uio.no}

\title[Optical properties TDOMP2]{Linear and nonlinear optical properties from TDOMP2 theory}

\begin{document}

\begin{abstract}
We present a derivation of the real-time time-dependent orbital-optimized Møller-Plesset (TDOMP2) theory
and its biorthogonal companion, time-dependent non-orthogonal OMP2 (TDNOMP2), theory starting from
the time-dependent bivariational principle and a parametrization based
on the exponential orbital-rotation operator formulation commonly used in 
time-independent molecular electronic-structure theory.
We apply the TDOMP2 method to extract absorption spectra and frequency-dependent
polarizabilities and first hyperpolarizabilities from real-time simulations, comparing the
results with those obtained from conventional time-dependent coupled-cluster singles and doubles (TDCCSD)
simulations and from its second-order approximation TDCC2. We also compare with
results from CCSD and CC2 linear and quadratic response theory. Our results indicate that while TDOMP2 absorption spectra are of the same quality as TDCC2 spectra, \added[comment={R1.1}]{including core excitations where optimized orbitals might be particularly important,}
frequency-dependent polarizabilities and hyperpolarizabilities from TDOMP2 simulations are significantly closer to TDCCSD results than those from TDCC2 simulations.
\end{abstract}

\section{Introduction}

The correct \added[comment={R2.1}]{semiclassical} description of interactions between matter and \added[comment={R2.1}]{temporally oscillating}
electromagnetic fields must start from time-dependent quantum mechanics.
Historically, the most-often used approach within molecular electronic-structure theory has been time-dependent perturbation theory where
the time-dependent Schr{\"o}dinger equation is solved order by order in the external field
strength,
leading to response theory
of molecular properties in the frequency domain through the application of a series of Fourier transforms.~\cite{Olsen1985}
Response theory has the advantage that it directly addresses the quantities that are used for the interpretation of experimental
measurements, such as one- and two-photon transition moments and frequency-dependent electric-dipole polarizabilities and
hyperpolarizabilities\added[comment={R2.1}]{, which may be expressed in terms of transition energies and stationary-state wave functions
that can, at least in principle, be obtained from the time-independent Schr{\"o}dinger equation for the particle system alone}.
A major disadvantage is that time resolution is lost when going from the time domain to the frequency domain. The obvious
solution would be to skip the Fourier transforms and instead work directly in the time domain. This, however, implies
that the time-dependent Schr{\"o}dinger equation must be solved order by order in a discretized time series, making the 
approach much too computationally demanding for higher-order properties. Instead, so-called \emph{real-time}
methods have received increasing attention in recent years---see, e.g., the review of real-time time-dependent electronic-structure theory
by \citeauthor{Li2020}~\cite{Li2020}

Real-time (RT) methods approximate the solution of the time-dependent Schr{\"o}dinger equation without
perturbation expansions and, thus, contain information about the response of the atomic or molecular electrons
to external electromagnetic fields to all orders in perturbation theory. Even extremely nonlinear processes that
are practically out of reach within response theory,
such as high harmonic generation and time-resolved one- and many-electron ionization probability amplitudes, are accessible
with RT methods, see Ref.~\citenum{Li2020} and references therein. Moreover, since RT methods include
the field explicitly in the simulation, it becomes possible to investigate the detailed dependence on laser
parameters such as intensity, frequency distribution, pulse shape, and delay between pump and probe pulses
without making explicit assumptions about the perturbation order of the electronic processes involved.

While RT methods are usually much simpler to implement than response theory (typically, the same code is needed
as for ground-state calculations, only generalized to complex parameters), a major downside of RT methods is the 
increased computational cost arising from the discretization of time. Thousands or even hundreds of thousands of
time steps are needed, each associated with a cost comparable to one (or a few) iterations of a ground-state
optimization with the same (time-independent) method. In addition, the basis-set requirements are generally
more demanding since, in principle, all excited states and even continuum states may be involved in the dynamics,
and acceleration techniques commonly used for ground-state and response calculations may not be generally applicable for
RT simulations with all possible external electromagnetic fields.

It is no surprise, therefore, that the most widely used RT electronic-structure method is real-time time-dependent
density-functional theory (RT-TDDFT)~\cite{Li2020,Runge1984,VanLeeuwen1999,Ullrich2012}.
Highly accurate wave function-based RT methods have also been developed, including multiconfigurational
time-dependent Hartree-Fock (MCTDHF)~\cite{Zanghellini2003,Kato2004,Meyer2009,Hochstuhl2014} theory
and related complete, restricted, and generalized active space formulations~\cite{Sato2013,Miyagi2013,Bauch2014}.
Avoiding the factorial computational scaling caused by the full configuration interaction (FCI) treatment at the heart
of these approaches, time-dependent extensions of
single-reference coupled-cluster (CC) theory~\cite{Bartlett2007}
and equation-of-motion CC (EOM-CC) theory~\cite{Krylov2008,Bartlett2012} have been increasingly often used to simulate laser-driven 
many-electron dynamics in the time domain in recent
years~\cite{Huber2011,OATDCC_2012,Nascimento2016,Nascimento2017,TDOCCD_Sato,Pedersen2019,Nascimento2019,Koulias2019,Bartlett_Core_2019,Pathak2020,OMP2_Sato_I,OMP2_Sato_II,Skeidsvoll2020,Kristiansen2020,Pedersen2021,Cooper2021,Park2021,Skeidsvoll2022}.
The two approaches, time-dependent CC (TDCC) and time-dependent EOM-CC (TD-EOM-CC) theory, differ in their parametrization
of the time-dependent left and right wave functions. While TDCC theory propagates the well-known exponential \emph{Ans{\"a}tze} for the wave functions,
TD-EOM-CC theory expresses them as linear combinations of EOM-CC left and right eigenstates. While both approaches are expected to give similar
results (and, indeed, appear to do so, see Ref.~\citenum{Skeidsvoll2022}) for weak-field processes, only TDCC theory (albeit with dynamical orbitals)
has been successfully applied to strong-field phenomena such as ionization dynamics and high harmonic generation~\cite{TDOCCD_Sato} to date.

Although the original formulation of TDCC theory in nuclear physics was based on time-dependent Hartree-Fock (HF) orbitals,~\cite{Hoodbhoy1978}
conventional TDCC theory is formulated with a static reference determinant, the HF ground state, which is kept fixed
during the dynamics in agreement with the conventional formulation of CC response (LRCC) theory~\cite{Koch1990,Pedersen1997}.
The fixed orbital space has some unwanted side effects, however. Gauge invariance is lost in truncated TDCC theory (but recovered in the
FCI limit)~\cite{Pedersen_OCC_1999,NOCC_Pedersen_2001}, severe numerical challenges arise as the CC ground state is depleted
during the dynamics (e.g., in ground--excited state Rabi oscillations)~\cite{Pedersen2019,Kristiansen2020}, and it becomes impossible to reduce
the computational effort whilst maintaining accuracy by splitting the orbital space into active and inactive orbitals for the
correlated treatment, as required to efficiently describe ionization dynamics~\cite{OATDCC_2012}. These deficiencies can, at least partially,
be circumvented by allowing the orbitals to move in concert with the electron correlation. In practice, this is done by
replacing the single excitations (and de-excitations) of conventional CC theory with full orbital rotations. This, in turn,
can be done in two ways. Within orbital-optimized CC (OCC) theory~\cite{Sherill_OCC_1998,Krylov1998,Pedersen_OCC_1999}, the orbitals
are required to remain orthonormal, whereas within nonorthogonal orbital-optimized CC (NOCC) theory~\cite{NOCC_Pedersen_2001,OATDCC_2012}
they are only required to be biorthonormal. The orthonormality constraint has an unfortunate side effect
in the sense that OCC theory does not converge to the FCI solution in
the limit of full rank cluster operators for three or more electrons, as pointed out by \citeauthor{Kohn_Olsen_OCC_not_FCI}~\cite{Kohn_Olsen_OCC_not_FCI}.
On the other hand, \citeauthor{Myhre_NOCC_FCI}~\cite{Myhre_NOCC_FCI} recently showed that NOCC theory may converge to the correct FCI limit
for any number of electrons.
In practice, however, time-dependent OCC (TDOCC) theory does not appear to deviate from the FCI limit by any significant amount~\cite{TDOCCD_Sato}.

The computational scaling with respect to the size of the basis set and with respect to the number of electrons
of TDOCC and time-dependent NOCC (TDNOCC) theory is essentially identical to that of conventional
TDCC theory with identical truncation of the cluster operators. The lowest-level truncation, after double excitations,
yields the TDOCCD and TDNOCCD methods that both scale as $\mathcal{O}(N^6)$, which is significantly more expensive than the
formal $\mathcal{O}(N^4)$ scaling of RT-TDDFT. In order to bring down the computational cost to a more tractable level,
\citeauthor{OMP2_Sato_I}~\cite{OMP2_Sato_I,OMP2_Sato_II} generalized the orbital-optimized second-order
M{\o}ller-Plesset (OMP2)~\cite{Bozkaya_OMP2_2011} method to the time domain and demonstrated 
that the resulting TDOMP2 method provides a reasonably accurate and gauge invariant description of highly nonlinear optical processes. 

In this work we assess the description of linear and quadratic optical properties within the TDOMP2 approximation. First, we review TDCC theory and its second-order approximation TDCC2. Second, we review time-dependent coupled-cluster theories with dynamic orbitals---TDNOCC and TDOCC theory---as obtained from the time-dependent bivariational principle, and introduce the second-order approximations TDNOMP2 and TDOMP2.
Finally, we compute linear (one-photon) absorption spectra and frequency-dependent polarizabilities and first hyperpolarizabilities with the TDOMP2, TDCCSD, and TDCC2 methods, and compare with results from CC2 and CCSD linear and quadratic response theory.

\section{Theory}

\subsection{Notation}
We consider a system of $N$ interacting electrons described by the second-quantized Hamiltonian
\begin{equation}
\hat{H} = \sum_{pq} h^p_q \hat{a}_p^\dagger \hat{a}_q + \frac{1}{2} \sum_{pqrs} u^{pq}_{rs} \hat{a}_p^\dagger \hat{a}_q^\dagger \hat{a}_s \hat{a}_r = \sum_{pq} h^p_q \hat{a}_p^\dagger \hat{a}_q + \frac{1}{4} \sum_{pqrs} v^{pq}_{rs} \hat{a}_p^\dagger \hat{a}_q^\dagger \hat{a}_s \hat{a}_r \label{Hamiltonian},
\end{equation}
where $\hat{a}_p^\dagger$ ($\hat{a}_p$) are creation (annihilation) operators associated with a finite set of $L$ orthonormal spin orbitals $\{ \phi_p \}_{p=1}^L$. The one- and two-body 
matrix elements $h^p_q$ and $u^{pq}_{rs}$ are defined as 
\begin{align}
h^p_q &= \braket{\phi_p|\hat{h}|\phi_q} = \int \phi_p^*(\mathbf{x}_1) \hat{h}(1) \phi_q(\mathbf{x}_1) \,d\mathbf{x}_1, \\
u^{pq}_{rs} &= \braket{\phi_p \phi_q|\hat{u}|\phi_r \phi_s} = \iint \phi_p^*(\mathbf{x}_1) \phi_q^*(\mathbf{x}_2) \hat{u}(1,2) \phi_r(\mathbf{x}_1) \phi_s(\mathbf{x}_2) \,d\mathbf{x}_1 \,d\mathbf{x}_2,
\end{align}
where $\mathbf{x}_i = (\mathbf{r}_i, \sigma_i)$ refers to the combined spatial-spin coordinate of electron $i$. The anti-symmetrized two-body matrix elements $v^{pq}_{rs}$ are given by
\begin{equation}
v^{pq}_{rs} = u^{pq}_{rs}-u^{pq}_{sr} \label{antisymmtbme}.
\end{equation}

\subsection{The TDCC2 approximation}
The TDCC \textit{Ansätze} for the left and right coupled-cluster wave functions are defined by
\begin{equation}
\ket{\Psi(t)} = e^{\hat{T}(t)}\ket{\Phi_0}, \qquad
\bra{\tilde{\Psi}(t)} = \bra{\Phi_0}\hat{\Lambda}(t)e^{-\hat{T}(t)},
\end{equation}
where $\ket{\Phi_0}$ is a reference determinant built from orthonormal spin orbitals, typically taken as the HF 
ground-state determinant. The chosen reference determinant splits the orbital set into occupied orbitals denoted by subscripts
$i,j,k,l$ and virtual orbitals denoted by subscripts $a,b,c,d$. Subscripts $p,q,r,s$ are used to denote general orbitals.
The cluster operators $\hat{T}(t)$ and $\hat{\Lambda}(t)$ are given by
\begin{align}
\hat{T}(t) &= \sum_\mu \tau^\mu(t) \hat{X}_\mu = \hat{T}_0 + \hat{T}_1 + \hat{T}_2 + \hat{T}_3 + \cdots + \hat{T}_N, \label{T-op}\\
\hat{\Lambda}(t) &= \sum_\mu \lambda_\mu(t) \hat{Y}^\mu = \hat{\Lambda}_0 + \hat{\Lambda}_1 + \hat{\Lambda}_2 + \hat{\Lambda}_3 + \cdots + \hat{\Lambda}_N \label{L-op},
\end{align}
where $\mu$ denotes excitations of rank $0,1,2,3,\ldots,N$,
and the excitation and de-excitation operators $\hat{X}_\mu$ and $\hat{Y}^\mu$ are defined by
\begin{align}
\label{Xdef}
\hat{X}_0 &\equiv 1, \qquad \hat{X}_\mu \ket{\Phi_0} \equiv \ket{\Phi_\mu}, \\
\label{Ydef}
\hat{Y}^0 &\equiv 1, \qquad \bra{\Phi_0} \hat{Y}^\mu \equiv \bra{\tilde{\Phi}_\mu},
\end{align}
such that $\braket{\tilde{\Phi}_\mu \vert \Phi_\nu} = \delta_{\mu\nu}$. The rank-$0$ cluster operators
are included to describe the phase and (intermediate) normalization of the CC state~\cite{Pedersen2019}.

The equations of motion for the wave function parameters are obtained from the bivariational action functional
used by \citeauthor{Arponen_1983}~\cite{Arponen_1983}
\begin{equation}
\mathcal{S}[\tilde{\Psi}, \Psi] = \int_{t_0}^{t_1} \mathcal{L} \,dt \label{action-functional},
\end{equation}
where the CC Lagrangian is given by
\begin{equation}
\mathcal{L} = \braket{\tilde{\Psi}(t)|\hat{H}(t)-i \partial_t|\Psi(t)} = \mathcal{H} -i\sum_\mu \lambda_\mu \dot{\tau}^\mu \label{CC-Lagrangian},
\end{equation}
and the Hamilton function $\mathcal{H}$ is given by
\begin{equation}
\mathcal{H} = \braket{\tilde{\Psi}(t)|\hat{H}(t)|\Psi(t)}.
\end{equation}
The requirement that $\mathcal{S}[\tilde{\Psi}, \Psi]$
be stationary with respect to variations of the complex parameters $z_\mu \in \{\tau^\mu, \lambda_\mu\}$ leads
to the Euler-Lagrange equations
\begin{equation}
\frac{\partial \mathcal{L}}{\partial z_\mu} = \frac{d}{dt} \frac{\partial \mathcal{L}}{\partial \dot{z}_\mu} \label{CC-Euler-Lagrange equations}.
\end{equation}
Taking the required derivatives yields the equations of motion for the amplitudes, 
\begin{align}
    i \dot{\tau}^\mu(t) &= \braket{\Phi_0|\hat{Y}^\mu e^{-\hat{T}(t)}\hat{H}(t)e^{\hat{T}(t)}|\Phi_0}, \label{cc-tau-eq-time} \\ 
    -i \dot{\lambda}_\mu(t) &= \braket{\Phi_0|\hat{\Lambda}(t) e^{-\hat{T}(t)}[\hat{H}(t), \hat{X}_\mu]e^{\hat{T}(t)}|\Phi_0} \label{cc-lambda-eq-time}.
\end{align}
Note that $\lambda_0(t)$ is a constant, which we choose such that the intermediate normalization condition $\braket{\tilde{\Psi}(t) \vert \Psi(t)} = 1$
is satisfied,
whereas the phase amplitude $\tau_0$ generally depends nontrivially on time~\cite{Pedersen2019}.
The phase amplitude may, however, be ignored as long as we are only interested in the time evolution of expectation values~\cite{Koch1990}.
For other quantities, such as the autocorrelation of the CC state~\cite{Pedersen2019} or certain stationary-state populations~\cite{Pedersen2021},
the phase amplitude is needed. In the present work, we will only consider expectation values.

Truncation of the cluster operators after single and double excitations defines the TDCCSD method, which has an asymptotic scaling of $\mathcal{O}(N^6)$. 
Defined as a second-order approximation to the TDCCSD method within many-body perturbation theory, the
TDCC2 method~\cite{Christiansen_cc2_1995} reduces the asymptotic scaling to $\mathcal{O}(N^5)$. In order to derive the TDCC2 equations, we partition the time-dependent Hamiltonian  
\begin{equation}
\hat{H}(t) = \hat{H}^{(0)}(t) + \hat{U}
\end{equation}
into a zeroth-order term,  $\hat{H}^{(0)}(t) = \hat{f}+\hat{V}(t)$, where $\hat{f}$ is the Fock operator,
and 
\begin{equation} 
\hat{V}(t) = \sum_{pq} (V)^p_q(t) \hat{a}_p^\dagger \hat{a}_q \label{td-oe-field-op}
\end{equation} 
is a time-dependent one-electron operator representing the interaction with an external field.
The first-order term (the fluctuation potential) is defined as, 
\begin{equation}
\hat{U} = \hat{H}(t)-\hat{f}-\hat{V}(t).
\end{equation}
In the many-body perturbation analysis of the TDCCSD equations,
the singles and doubles amplitudes are considered zeroth-order and first-order quantities, respectively.
For notational convenience the time-dependence of the amplitudes and operators will be understood implicitly in the following.

Equations of motion are obtained from making the action given by Eq.~(\ref{action-functional}) stationary with respect to variations of the amplitudes.
The TDCC2 Lagrangian is obtained from the TDCCSD Lagrangian by retaining terms up to quadratic in the doubles amplitudes and the fluctuation potential, 
\begin{equation}
\mathcal{L} = \mathcal{H} - i \left( \sum_{\mu_1} \lambda_{\mu_1} \dot{\tau}^{\mu_1} + \sum_{\mu_2} \lambda_{\mu_2} \dot{\tau}^{\mu_2} \right).
\end{equation}
Introducing $\hat{T}_1$-transformed operators as 
\begin{equation}
\tilde{\Omega} = e^{-\hat{T}_1} \hat{\Omega} e^{\hat{T}_1},
\end{equation}
the TDCC2 approximation to the TDCCSD Hamilton function becomes
\begin{align}
\mathcal{H} &= \bra{\Phi_0} e^{-\hat{T}_2}  \tilde{H} e^{\hat{T}_2} \ket{\Phi_{0}} \nonumber \\
            &+ \sum_{\mu_1} \lambda_{\mu_1} \bra{\tilde{\Phi}_{\mu_1}} \tilde{H}+[\tilde{H}, \hat{T}_2] \ket{\Phi_{0}} 
             + \sum_{\mu_2} \lambda_{\mu_2} \bra{\tilde{\Phi}_{\mu_2}} [\hat{f} +\tilde{V}, \hat{T}_2] + \tilde{U}  \ket{\Phi_{0}}. 
\label{eq:HCC2}
\end{align}
Note that the Fock operator appearing in the commutator in the last term is \textit{not} $\hat{T}_1$ transformed. 
The Euler-Lagrange equations then yield
equations of motion for the singles amplitudes,
\begin{align}
i \dot{\lambda}^i_a &= (f_1)^{i}_{a}           
+  (f_1)^{b}_{a} \lambda^{i}_{b}        
- (f_1)^{i}_{j} \lambda^{j}_{a}       
 + \lambda^{j}_{b} \tilde{v}^{ib}_{aj}          
+ \frac{1}{2}\sum_c \lambda^{ij}_{bc} \tilde{v}^{bc}_{aj}   \nonumber
- \frac{1}{2} \sum_k \lambda^{jk}_{ab} \tilde{v}^{ib}_{jk}   \\ 
&+ \sum_{ck}\left( \lambda^{j}_{b} \tau^{bc}_{jk} \tilde{v}^{ik}_{ac}    
- \frac{1}{2} \lambda^{i}_{b} \tau^{bc}_{jk} \tilde{v}^{jk}_{ac}   
- \frac{1}{2} \lambda^{j}_{a} \tau^{bc}_{jk} \tilde{v}^{ik}_{bc} \right),   \\
i \dot{\tau}^a_i &= (f_1)^a_i     
   + \sum_{jb} (f_1)^j_b \tau_{ji}^{ba}   
   + \frac{1}{2} \sum_{jbc} \tau_{ik}^{bc} \tilde{v}^{aj}_{bc} 
   +\frac{1}{2} \sum_{jkc} \tau_{jk}^{ac} \tilde{v}^{jk}_{bi},
\end{align}
and for the doubles amplitudes,
\begin{align}
i \dot{\tau}^{ab}_{ij}   &= \tilde{v}^{ij}_{ab}  
+  P(ab) \sum_{c} (f_2)^a_c \tau_{ij}^{cb} + P(ij)\sum_{k} (f_2)^k_j \tau_{ki}^{ab}, \\  
i \dot{\lambda}^{ij}_{ab}  &=
\tilde{v}^{ij}_{ab}
+ \hat{P}(ab) \hat{P}(ij)(f_1)^{i}_{a} \lambda^{j}_{b} - \sum_c \left( \hat{P}(ab)(f_2)^{c}_{a} \lambda^{ij}_{bc}  
+ \hat{P}(ij) \lambda^{i}_{c} \tilde{v}^{jc}_{ab} \right) \nonumber \\
&+\sum_k \left( \hat{P}(ij) (f_2)^{i}_{k} \lambda^{jk}_{ab}+\hat{P}(ab) \lambda^{k}_{a} \tilde{v}^{ij}_{bk} \right).
\end{align}
Here, we have defined the fully and partially $T_1$-transformed Fock matrices
\begin{equation}
(f_1)^p_q \equiv \tilde{f}^p_q + (\tilde{V})^p_q, \qquad
(f_2)^p_q \equiv f^p_q + (\tilde{V})^p_q,
\end{equation}
and the operator $\hat{P}(pq)$ is an anti-symmetrizer defined by its action on the elements of an arbitray tensor $M$:  
\replaced[comment={R2.2}]{$\hat{P}(pq) M_{pq} = M_{pq} - M_{qp}$.}{$\hat{P}(pq) M^p_q = M^p_q - M^q_p$.}

\added[comment={R2.7}]{The presence of the untransformed Fock operator in Eq.~\eqref{eq:HCC2} has a number of
simplifying consequences. For example, the ground-state doubles amplitudes become explicit functions of the singles amplitudes
and the double excitation block of the EOM-CC Hamiltonian matrix (the CC Jacobian) becomes diagonal. In TDCC2 theory, however,
it implies that the doubles amplitudes are not fully adjusted to the approximate orbital relaxation captured by the 
(zeroth order) singles amplitudes. In order to test the consequences of this, we have implemented the
TDCC2-b method of \citeauthor{kats2006local_cc2b}~\cite{kats2006local_cc2b}, where the fully $T_1$-transformed Fock operator
is used in Eq.~\eqref{eq:HCC2}.}

\subsection{Review of time-dependent coupled-cluster theories with dynamic orbitals}

The TDOCC and TDNOCC \textit{Ansätze} replace the singles amplitudes of conventional TDCC theory with unitary and non-unitary orbital rotations, respectively.
For both types of orbital rotations, the left and right coupled-cluster wave functions can be written on the form
\begin{equation}
\ket{\Psi(t)} = e^{\hat{\kappa}(t)}e^{\hat{T}(t)}\ket{\Phi_0}, \qquad
\bra{\tilde{\Psi}(t)} = \bra{\Phi_0}\hat{\Lambda}(t)e^{-\hat{T}(t)}e^{-\hat{\kappa}(t)},
\end{equation}
where $\ket{\Phi_0}$ is a static reference determinant built from orthonormal spin orbitals, typically taken as the HF
ground-state determinant in analogy with conventional TDCC theory.
The terminology of occupied and virtual orbitals thus refers to this reference determinant, although both subsets are changed by the
time-dependent orbital rotations.
Excluding singles amplitudes, the cluster operators $\hat{T}(t)$ and $\hat{\Lambda}(t)$ are given by
\begin{align}
\hat{T}(t) &= \sum_\mu \tau^\mu(t) \hat{X}_\mu = \hat{T}_0 + \hat{T}_2 + \hat{T}_3 + \cdots + \hat{T}_N, \\
\hat{\Lambda}(t) &= \sum_\mu \lambda_\mu(t) \hat{Y}^\mu = \hat{\Lambda}_0 + \hat{\Lambda}_2 + \hat{\Lambda}_3 + \cdots + \hat{\Lambda}_N,
\end{align}
where $\mu$ denotes excitations of rank $0,2,3,\ldots,N$,
and the excitation and de-excitation operators $\hat{X}_\mu$ and $\hat{Y}^\mu$ are defined the same way as in conventional
TDCC theory [Eqs.~\eqref{Xdef} and \eqref{Ydef}]. 
The exclusion of singles amplitudes is rigorously justified, as they become 
redundant when the orbitals are properly relaxed by the orbital-rotation operator $\exp(\hat{\kappa})$~\cite{Pedersen_OCC_1999,NOCC_Pedersen_2001,OATDCC_2012}.

In TDNOCC theory, the orbital rotations are non-unitary, i.e. $\hat{\kappa}^\dagger \neq -\hat{\kappa}$. If $\hat{\kappa}$ is restricted to be anti-Hermitian, we obtain TDOCC theory where the orbital rotations are unitary. However, this leads to the parametrization formally not converging to the FCI limit (for $N>2$), as pointed out by \citeauthor{Kohn_Olsen_OCC_not_FCI}~\cite{Kohn_Olsen_OCC_not_FCI}. On the other hand, \citeauthor{Myhre_NOCC_FCI}~\cite{Myhre_NOCC_FCI} showed that the proper FCI limit may be restored by non-unitary orbital rotations. Furthermore, it can be shown that occupied-occupied and virtual-virtual rotations are redundant~\cite{Pedersen_OCC_1999,OATDCC_2012} and it is sufficient to consider $\hat{\kappa}(t)$ on the form 
\begin{equation}
\hat{\kappa}(t) = \sum_{ai} \left( \kappa^a_i(t) \hat{X}^a_i + \kappa^i_a(t) \hat{Y}^i_a \right).
\end{equation}

Using the Baker-Campbell-Hausdorff expansion, one can show that the similarity transforms of the creation and annihilation operators with $\exp(\hat{\kappa})$ are given by
\begin{align}
e^{-\hat{\kappa}(t)}\hat{a}_p^\dagger e^{\hat{\kappa}(t)} &= \sum_q \hat{a}_q^\dagger [e^{-\kappa(t)}]^q_p, \\
e^{-\hat{\kappa}(t)}\hat{a}_p e^{\hat{\kappa}(t)} &= \sum_q \hat{a}_q [e^{\kappa(t)}]^p_q.
\end{align}
Recalling that explicit time-dependence only appears in the interaction operator and in the wave function parameters, we will suppress 
the dependence on time in the notation.
For a general one- and two-body operator $\hat{\Omega}$,
the TDNOCC and TDOCC expectation value functionals can be written as
\begin{equation}
\braket{\tilde{\Psi}(t)|\hat{\Omega}|\Psi(t)} = \sum_{pq}\tilde{\Omega}^p_q \gamma^q_p + \frac{1}{4}\sum_{pqrs} \tilde{\Omega}^{pq}_{rs}\Gamma^{rs}_{pq},
\end{equation}
where 
\begin{align}
    \tilde{\Omega}^p_q &= \sum_{r s} [e^{-\kappa}]^p_r \Omega^r_s [e^{\kappa}]^s_q, \label{kappa-trans-obop}\\
    \tilde{\Omega}^{pq}_{rs} &= \sum_{t u v w} [e^{-\kappa}]^p_t [e^{-\kappa}]^q_u \Omega^{tu}_{vw} [e^{\kappa}]^v_r [e^{\kappa}]^w_s \label{kappa-trans-tbop},
\end{align}
and $\gamma, \Gamma$ are effective one- and two-body density matrices given by
\begin{align}
\gamma^q_p  &= \braket{\Phi_0|\hat{\Lambda}(t)e^{-\hat{T}(t)} \hat{a}_p^\dagger \hat{a}_qe^{\hat{T}(t)}|\Phi_0}, \\
\Gamma_{pq}^{rs}  &= \braket{\Phi_0|\hat{\Lambda}(t)e^{-\hat{T}(t)} \hat{a}_p^\dagger \hat{a}_q^\dagger \hat{a}_s \hat{a}_re^{\hat{T}(t)}|\Phi_0}.
\end{align}

The equations of motion for the wave function parameters are, again, obtained from the Euler-Lagrange equations~\eqref{CC-Euler-Lagrange equations}
for the full parameter set $z_\mu \in \{\tau^\mu, \lambda_\mu, \kappa^a_i, \kappa^i_a \}$
with the Lagrangian given by
\begin{align}
\mathcal{L} &= \braket{\tilde{\Psi}|\hat{H}-i \partial_t|\Psi} \nonumber \\ 
&= \mathcal{H} - i \braket{\tilde{\Psi}|\hat{Q}_1|\Psi} -i\sum_\mu \lambda_\mu \dot{\tau}^\mu \label{NOCC-Lagrangian},
\end{align}
\added[comment={R2.4}]{where the Hamiltonian is given by Eq.~(\ref{Hamiltonian}). Here, the interaction with the external field~(\ref{td-oe-field-op}) is absorbed into the one-body part of the Hamiltonian such that
\begin{equation}
h^p_q \leftarrow h^p_q + (V)^p_q(t).
\end{equation}}
The operator $\hat{Q}_1$ is defined as
\begin{equation}
\hat{Q}_1 \equiv \frac{\partial e^{\hat{\kappa}}}{\partial t}e^{-\hat{\kappa}},
\end{equation}
and $\mathcal{H} = \braket{\tilde{\Psi}|\hat{H}\added{(t)}|\Psi}$.

The detailed derivation of the equations of motion is greatly simplified by absorbing the orbital rotation in the Hamiltonian at each point in
time,
$\hat{H} \leftarrow \exp(-\hat{\kappa})\hat{H}\exp(\hat{\kappa})$, which amounts to temporally local updates of the Hamiltonian integrals
according to Eqs.~\eqref{kappa-trans-obop} and \eqref{kappa-trans-tbop}.
This allows us to compute the temporally local derivatives of the Lagrangian with respect to the parameters at
the point $\hat{\kappa} = 0$ such that, for example, the rather complicated operator $\hat{Q}_1$ becomes the much simpler
operator $\hat{\dot{\kappa}}$.
We thus find that the equations of motion for the cluster amplitudes are given by
\begin{align}
i \dot{\tau}^\mu &= \braket{\Phi_0|\hat{Y}^\mu e^{-\hat{T}}\left(\hat{H}-i\hat{\dot{\kappa}}\right)e^{\hat{T}}|\Phi_0}, \label{tau-eq}  \\
-i\dot{\lambda}_\mu &= \braket{\Phi_0|\hat{\Lambda}e^{-\hat{T}}\left[\left(\hat{H}-i\hat{\dot{\kappa}}\right), \hat{X}_\mu \right]e^{\hat{T}}|\Phi_0}, \label{lambda-eq}
\end{align}
where the right-hand sides are essentially identical to the usual amplitude equations of coupled-cluster theory with additional terms
arising from the one-body operator $\hat{\dot{\kappa}}$. As in conventional TDCC theory, $\lambda_0$ is constant and may be chosen such that
intermediate normalization is preserved~\cite{Pedersen2019}.
In the same manner, we may derive the equations of motion for the orbital-rotation parameters as
\begin{align}
i\sum_{bj} \dot{\kappa}^j_b A^{ib}_{aj} = R^i_a, \label{kappa-down-eq}\\ 
-i\sum_{bj} \dot{\kappa}^b_j A^{ja}_{bi} = R^a_i \label{kappa-up-eq},
\end{align}
where the right-hand sides are given by Eqs. (30a) and (30b) in Ref.~\citenum{OATDCC_2012}, and 
\begin{equation}
A^{ib}_{aj} = \braket{\tilde{\Psi}|[\hat{a}_j^\dagger \hat{a}_b, \hat{a}_a^\dagger \hat{a}_i]|\Psi} = \delta^b_a \gamma^i_j-\delta^i_j \gamma^b_a \label{A-tensor}.
\end{equation}
Equations (\ref{kappa-down-eq}) and (\ref{kappa-up-eq}) are linear systems of algebraic equations which require the matrix $A = [A^{ib}_{aj}]$ to be non-singular in order to have a unique solution. 
We remark that this matrix becomes singular whenever an eigenvalue of the occupied density block is equal to an eigenvalue of the virtual density block.
While this would prevent straightforward integration of the orbital equations of motion, we have not encountered the singularity in actual simulations
thus far.

The above derivation does not require unitary orbital rotations and is, therefore, applicable to TDNOCC theory. Specialization to TDOCC theory is
most conveniently done by starting from the inherently real
action functional~\cite{Pedersen_OCC_1999,TDOCCD_Sato} 
\begin{equation}
\mathcal{S} = \Re\int_{t_0}^{t_1} \mathcal{L} \,dt = \int_{t_0}^{t_1}\frac{1}{2}\left(\mathcal{L}+\mathcal{L}^*\right) \,dt \label{TDOCC-action},
\end{equation} 
which is required stationary with respect to variations of all parameters. The expression for the Lagrangian $\mathcal{L}$ is identical to
Eq. (\ref{NOCC-Lagrangian}) with $\hat{\kappa}$ anti-Hermitian. 
The Euler-Lagrange equations then take the form
\begin{equation}
0 = \frac{1}{2} \left(\frac{\partial \mathcal{L}}{\partial z_\mu}-\frac{d}{dt}\frac{\partial \mathcal{L}}{\partial \dot{z}_\mu}\right)+\frac{1}{2} \left(\frac{\partial \mathcal{L}}{\partial z_\mu^*}-\frac{d}{dt}\frac{\partial \mathcal{L}}{\partial \dot{z}_\mu^*}\right)^* \label{Euler-Lagrange-OCC},
\end{equation}
for $z_\mu \in \{\kappa^i_a, \lambda_\mu, \tau^\mu\}$. 
The derivatives of $\mathcal{L}$ with respect to the complex-conjugated parameters vanish for the amplitudes $\lambda_\mu$ and $\tau^\mu$ and, therefore,
the resulting equations of motion for the amplitudes are identical to Eqs.~(\ref{tau-eq}) and (\ref{lambda-eq}).

Taking the derivative of $\mathcal{L}$ with respect to $\kappa^a_i$ and using Eqs.~(\ref{kappa-down-eq})-(\ref{A-tensor}) we obtain the equations of motion for the orbital-rotation parameters  
\begin{align}
i\sum_{bj} \dot{\kappa}^j_b B^{ib}_{aj} = \sum_p h^p_a D^i_p - \sum_q h^i_q D^q_a + \frac{1}{2} \left( \sum_{pqr}v^{pq}_{ra} P^{ri}_{pq} - \sum_{qrs} v^{iq}_{rs} P^{rs}_{aq} \right) + i\dot{D}^i_a,
\end{align}
where we have defined the hermitized one- and two-body density matrices
\begin{align}
D^p_q &= \frac{1}{2} \left(\gamma^p_q + \gamma_p^{q*} \right), \\
P^{pq}_{rs} &= \frac{1}{2} \left( \Gamma^{pq}_{rs} + \Gamma^{rs*}_{pq} \right)
\end{align}
and the matrix
\begin{equation}
B^{ib}_{aj} = \delta^b_a D^i_j - \delta^i_j D^b_a.
\end{equation}
Here, too, we face a potential singularity, which we have never encountered in practical simulations thus far.

\subsection{TDOMP2 theory}
In the spirit of the TDCC2 approximation to TDCCSD theory, we may introduce second-order approximations to TDNOCCD and TDOCCD theory, which we
will designate TDNOMP2 and TDOMP2 theory, respectively, in accordance with the naming convention used in time-independent theory~\cite{Bozkaya_OMP2_2011}.
The TDOMP2~\cite{OMP2_Sato_I, OMP2_Sato_II} method has previously been formulated as a second-order approximation to the
TDOCCD method~\cite{TDOCCD_Sato, Pedersen_OCC_1999} by \citeauthor{OMP2_Sato_I}~\cite{OMP2_Sato_I, OMP2_Sato_II}
The definition of perturbation order is analogous to that of the TDCC2 approximation to the TDCCSD method~\cite{Christiansen_cc2_1995}, as outlined above.
Thus, the Hamiltonian is split into a zeroth-order term, \replaced[comment={R2.4}]{$\hat{H}^{(0)}(t) = \hat{f}+\hat{V}(t)$, and a first-order term, the fluctuation potential $\hat{U} = \hat{H}(t)-\hat{f}-\hat{V}(t)$
such that the HF reference determinant is the ground state of the zeroth order Hamiltonian for $\hat{V}(t) \rightarrow 0$.}{the Fock operator $\hat{f}$, and a first-order term, the fluctuation potential $\hat{H}-\hat{f}$
such that the HF reference determinant is the zeroth-order ground-state wave function.}
The doubles amplitudes enter at the first-order level, whereas the orbital-rotation parameters are considered zeroth order in analogy with the singles
amplitudes of TDCC2 theory~\cite{Christiansen_cc2_1995}.
 
We start by considering non-unitary orbital rotations and introduce the
$\hat{\kappa}$-transformed operators
\begin{equation}
\tilde{\Omega} = e^{-\hat{\kappa}} \hat{\Omega} e^{\hat{\kappa}}.
\end{equation} 
The TDNOMP2 Lagrangian is defined by truncating the cluster operators at the doubles level and retaining terms up to quadratic 
in $(\lambda, \tau, u)$ in the TDNOCC Lagrangian (\ref{NOCC-Lagrangian}),
\begin{align}
\mathcal{L} = \mathcal{H}- i\sum_{abij}\lambda^{ij}_{ab}\dot{\tau}^{ab}_{ij} -i \braket{\Phi_0|\left(1+\hat{\Lambda}_2\right)\left(\tilde{Q}_1+[\tilde{Q}_1,\hat{T}_2]\right)|\Phi_0} \label{NOMP2-Lagrangian}.
\end{align}
The TDNOMP2 Hamilton function $\mathcal{H}$ becomes
\begin{align}
\mathcal{H} = \braket{\Phi_0|\tilde{H}+[\tilde{H}, \hat{T}_2]+\hat{\Lambda}_2\tilde{H}+\hat{\Lambda}_2[\tilde{F}, \hat{T}_2]|\Phi_0} = \sum_{pq} \tilde{h}^p_q \gamma_p^q + \frac{1}{4} \sum_{pqrs} \tilde{v}^{pq}_{rs} \Gamma_{pq}^{rs},
\end{align}
where $\tilde{h}^p_q, \tilde{v}^{pq}_{rs}$ are matrix elements transformed according to Eqs. (\ref{kappa-trans-obop}) and (\ref{kappa-trans-tbop}). 
\added[comment={R2.6}]{The operator $\tilde{F}$ is given by
\begin{equation}
\tilde{F} = \sum_{pq} \tilde{f}^p_q \hat{a}_p^\dagger \hat{a}_q
\end{equation}
where 
\begin{equation}
\tilde{f}^p_q = \braket{\Phi_0|[\hat{a}_q^\dagger, [\hat{a}_p, \tilde{H}]]_+|\Phi_0}
              = \tilde{h}^p_q + \sum_j \tilde{v}^{pj}_{qj}.
\end{equation}
}

The non-zero matrix elements of the TDNOMP2 one- and two-body density matrices $\gamma, \Gamma$ are given by
\begin{align}
\gamma_i^j &= \delta_i^j+ (\gamma_{\text{c}})^j_i, \ \ (\gamma_{\text{c}})^j_i = -\frac{1}{2} \sum_k \lambda^{jk}_{ab}\tau^{ab}_{ik}, \ \ \gamma^b_a = \frac{1}{2} \sum_c \lambda^{ij}_{ac}\tau^{bc}_{ij}, \label{onerdm-oo-onerdm-vv}\\
\Gamma^{kl}_{ij} &= \delta^k_i\delta^l_j - \delta^k_j \delta^l_i + \hat{P}(kl)\hat{P}(ij) \delta^k_i (\gamma_{\text{c}})^l_j, \label{twordm-oooo}\\
\Gamma^{ab}_{ij} &= \tau^{ab}_{ij}, \ \ \Gamma^{ij}_{ab} = \lambda^{ij}_{ab}, \label{twordm-vvoo-twordm-oovv}\\
\Gamma^{bj}_{ak} &= -\Gamma^{jb}_{ak} = -\Gamma^{bj}_{ka} = \Gamma^{jb}_{ka} = \delta^j_k \gamma_a^b. \label{twordm-ovvo}
\end{align}

Equations of motion now follow from the Euler-Lagrange equations with the Lagrangian given by Eq. (\ref{NOMP2-Lagrangian}). Taking the required derivatives and the $\hat{\kappa} \rightarrow 0$ limit 
we find the equations of motion for the amplitudes
\begin{align}
i\dot{\tau}^{ab}_{ij} &= v^{ab}_{ij} - \hat{P}(ij) \sum_k f^k_j\tau^{ab}_{ik} + \hat{P}(ab) \sum_c f^a_c \tau^{cb}_{ij}, \label{tau-nomp2}\\
- i \dot{\lambda}^{ij}_{ab} &= v^{ij}_{ab} - \hat{P}(ij) \sum_k f^i_k \lambda^{kj}_{ab} + \hat{P}(ab) \sum_c f^c_a \lambda^{ij}_{cb} \label{lambda-nomp2}.
\end{align}
The time-dependence of the orbital-rotation parameters in the $\hat{\kappa} \rightarrow 0$ limit takes the same form as Eqs. (\ref{kappa-down-eq}) and (\ref{kappa-up-eq}) with density matrices given by Eqs. (\ref{onerdm-oo-onerdm-vv})-(\ref{twordm-ovvo}). Explicit insertion of non-zero matrix elements yields
\begin{align}
i\sum_{bj} \dot{\kappa}^j_b A^{ib}_{aj} &= \sum_j f^j_a \gamma^i_j - \sum_b f^i_b \gamma^b_a
                                         + \sum_{jl} v^{il}_{aj}(\gamma_c)^j_l+\sum_{bc}v^{ib}_{ac}\gamma^c_b \nonumber \\
&+ \frac{1}{2} \left( \sum_{jbc}v^{bc}_{aj}\lambda^{ij}_{bc} - \sum_{klc} v^{ic}_{kl} \lambda^{kl}_{ac}\right), \\
i\sum_{bj} \dot{\kappa}^b_j A^{aj}_{ib} &= \sum_b f^b_i \gamma^a_b - \sum_j f^a_j \gamma^j_i
                                         - \sum_{jl}v^{aj}_{il}(\gamma_c)^l_j-\sum_{bc}v^{ac}_{ib}\gamma^b_c \nonumber \\
&+ \frac{1}{2} \left( \sum_{klc} v^{kl}_{ic} \tau^{ac}_{kl} - \sum_{jbc}v^{aj}_{bc}\tau^{bc}_{ij} \right).
\end{align}

We can now obtain the TDOMP2 equations from the TDNOMP2 equations. The action functional takes the form of Eq. (\ref{TDOCC-action}) where $\mathcal{L}$ is equivalent to the expression given by Eq. (\ref{NOMP2-Lagrangian}) with $\hat{\kappa} = -\hat{\kappa}^\dagger$ and equations of motion are obtained from the Euler-Lagrange equation (\ref{Euler-Lagrange-OCC}).
Since the derivatives of the Lagrangian with respect to the complex-conjugated amplitudes are zero, equations of motion for the amplitudes are equivalent to Eqs. (\ref{tau-nomp2}) and (\ref{lambda-nomp2}).
However, since the orbital transformation is orthonormal, $h, u$ and $f$ are Hermitian and it follows that the equation for $\lambda^{ij}_{ab}$ is just the complex-conjugate of that for $\tau^{ab}_{ij}$ such that 
\begin{equation}
\lambda^{ij}_{ab} = \tau^{ab*}_{ij}
\end{equation}
and, thus, it is sufficient to solve only one of the two sets of amplitude equations.
This simplification arises from the unitarity of the orbital rotations and is not obtained within neither TDCC2 nor TDNOMP2 theory.
In addition, it follows that the one- and two-body density matrices given by eqs. (\ref{onerdm-oo-onerdm-vv})-(\ref{twordm-ovvo}) are Hermitian, i.e., 
\begin{equation}
\gamma^p_q = \gamma_{p}^{q*}, \qquad
\Gamma^{pq}_{rs} = \Gamma_{pq}^{rs*}.
\end{equation}

From the Euler-Lagrange equation we then find that the equation of motion for $\kappa^i_a$ is given by, 
\begin{align}
i\sum_{bj} \dot{\kappa}^j_b A^{ib}_{aj} &= \sum_j f^j_a \gamma^i_j - \sum_b f^i_b \gamma^b_a
                                         + \sum_{jl} v^{il}_{aj}(\gamma_c)^j_l+\sum_{bc}v^{ib}_{ac}\gamma^c_b \nonumber \\
&+ \frac{1}{2} \left( \sum_{jbc}v^{bc}_{aj}(\tau^{bc}_{ij})^* - \sum_{klc} v^{ic}_{kl} (\tau^{ac}_{kl})^*\right).
\end{align}
Note that in contrast to the TDOCC equations there is no need to explicitly hermitize the density matrices, as they already are Hermitian within TDOMP2 theory.
     
\subsection{Optical properties from real-time simulations}
\label{sec:opt_props_from_rtsims}
In order to extract linear and nonlinear optical properties from real-time time-dependent simulations we subject an atom or molecule, initially in its (electronic) ground state, to a time-dependent electric field $\mathcal{E}(t)$.
The semiclassical interaction operator in the electric-dipole approximation in the length gauge is given by
\begin{equation}
\hat{V}(t) = -\sum_{i \in \{x,y,z\}} \mu_i \mathcal{E}_i(t),
\end{equation}
where $\mu_i$ is the $i$th Cartesian component of the electric dipole moment operator.
The shape, frequency, and strength of the electric field determines which properties may be extracted from time-dependent simulations.

Linear (one-photon) absorption spectra can be computed by using a weak electric-field impulse to induce transitions from the electronic ground state
to all electric-dipole allowed excited states of the system~\cite{Repsisky_2015, goings2016atomic}, including core excitations as well as valence excitations.
Such an electric-field kick is represented by
the delta pulse $\mathcal{E}(t) = \mathcal{E}_\text{max}\delta(t)$, which we discretize by means of the box function
\begin{equation}
\mathcal{E}_i(t) = 
\left\{
\begin{array}{ll}
\mathcal{E}_{\text{max}} n_i & \qquad 0 \leq t  < \Delta t, \\
0 & \qquad \text{else},
\end{array}
\right.
\label{delta_pulse}
\end{equation}
where $\mathcal{E}_{\text{max}}$ is the strength of the field,
$n_{i}$ is the $i$th Cartesian component of the real unit polarization vector $\vec{n}$, and
$\Delta t$ is the time step of the simulation.

The absorption spectrum is computed from the relation
\begin{equation}
S(\omega) = \frac{4\pi \omega}{3c} \text{Im} \text{Tr}[\mathbf{\alpha}(\omega)],
\end{equation}
where the frequency-dependent dipole polarizability tensor $\mathbf{\alpha}(\omega)$ is obtained from the Fourier transform of the
induced dipole moment
\begin{equation}
\mu_{ij}^{\text{ind}}(t) = \mu_{ij}(t)-\mu_i^0.
\end{equation}
Here, $\mu_{ij}^{\text{ind}}(t)$ is the $i$th component of the induced dipole moment with the field polarized in the direction $j \in \{x,y,z\}$,
$\mu_i^0$ is the $i$th component of the permanent dipole moment, and $\mu_{ij}(t)$ is computed as the trace of the dipole matrix in the orbital
basis and the effective one-body density matrix (in the same basis).
In practice,
we only compute finite signals at discrete points in time, forcing us to use the Fast Fourier Transform algorithm (FFT).
In order to avoid artefacts arising from the periodicity of the FFT, we premultiply the dipole signal with the exponential damping factor $\exp(-\gamma t)$,
\begin{equation}
\mathbf{\alpha}_{ij}(\omega) = \text{FFT}(\mu^{\text{ind}}_{ij}(t) e^{-\gamma t})/\mathcal{E}_{\text{max}}, \label{fourier_lorentz} 
\end{equation}
where $\gamma > 0$ is chosen such that the induced dipole moment vanishes at the end of the simulation.
This choice of damping factor artificially broadens the excited energy levels, producing Lorentzian line shapes in 
the computed spectra.

Also dynamic polarizabilities and hyperpolarizabilities can be extracted from real-time time-dependent simulations using the method
described by~\citeauthor{Li_hyperpol_2013}~\cite{Li_hyperpol_2013}
Suppose that the system under consideration interacts with a weak adiabatically switched-on monochromatic electric field,
\begin{equation}
\mathcal{E}(t) = \mathcal{E}_0 \cos(\omega t),
\end{equation}
where $\omega$ is the frequency and $\mathcal{E}_0$ is the amplitude of the field. The dipole moment can then be written as a series expansion in the electric
field strength,
\begin{equation}
\mu_i(t) = \mu_i^0 + \sum_{j \in \{x,y,z\}} \mu_{ij}^{(1)}(t) \mathcal{E}_j + \sum_{j,k \in \{x,y,z\}} \mu_{ijk}^{(2)}(t)\mathcal{E}_j \mathcal{E}_k + \cdots, 
\end{equation} 
provided that $\mathcal{E}_0$ is sufficiently small and $\omega$ belongs to a transparent spectral region of the system at hand.
The time-dependent dipole response functions $\mu_{ij}^{(1)}(t)$ and $\mu_{ijk}^{(2)}(t)$ can be expressed as
\begin{align}
\mu_{ij}^{(1)}(t) &= \alpha_{ij}(-\omega; \omega) \cos(\omega t) \label{polarizability}, \\
\mu_{ijk}^{(2)}(t) &= \frac{1}{4} \left[ \beta_{ijk}(-2\omega; \omega, \omega) \cos(2\omega t) + \beta_{ijk}(0; \omega, -\omega) \right], \label{first-polarizability}
\end{align}
where $\alpha_{ij}, \beta_{ijk}$ are Cartesian components of the polarizability and first hyperpolarizability tensors.
The ``diagonal'' elements $\mu_{ij}^{(1)}, \mu_{ijj}^{(2)}$ of the dipole response functions
can be calculated from the time-dependent signal using the four-point central difference formulas,
\begin{align}
\mu_{ij}^{(1)} &\approx \frac{8[\mu_i(t, \mathcal{E}_j)-\mu_i(t, -\mathcal{E}_j)]-[\mu_i(t, 2\mathcal{E}_j)-\mu_i(t, -2\mathcal{E}_j)]}{12 \mathcal{E}_j}, \label{first_order_dip_response}\\
\mu_{ijj}^{(2)} &\approx 
	\frac{16[\mu_i(t, \mathcal{E}_j) + \mu_i(t, -\mathcal{E}_j)] 
	- [\mu_i(t, 2\mathcal{E}_j) + \mu_i(t, -2\mathcal{E}_j)] 
	- 30\mu_i^{0}}
	{24 \mathcal{E}_j^2}, \label{second_order_dip_response}
\end{align}
with $\mu_i(t, \mathcal{E}_j)$ being the $i$th component of the time-dependent dipole moment when a cosine field with strength of $\mathcal{E}_j$ in the $\pm j$th direction is applied. Finally, 
the polarizabilities and first hyperpolarizabilities are determined by performing a curve fit of the dipole response functions computed with finite differences 
to the analytical forms given by Eqs. (\ref{polarizability}) and (\ref{first-polarizability}). 

In practice, it is infeasible to adiabatically switch on the electric field. This is circumvented by~\citeauthor{Li_hyperpol_2013}~\cite{Li_hyperpol_2013}
by turning on the field with a linear ramping envelope lasting for one optical cycle,
\begin{equation}
t_c = \frac{2 \pi }{\omega}.
\end{equation}
The electric field is then given by
\begin{equation}
\mathcal{E}(t) = 
\left\{
\begin{array}{ll}
   \frac{t}{t_c} \mathcal{E}_0 \cos(\omega t) & \qquad 0 \leq t < t_c, \\
   \mathcal{E}_0\cos(\omega t)  & \qquad t \geq t_c,
\end{array}
\right. \label{ramp_field}
\end{equation}  
and the curve fit is performed only on the part of the signal computed after the ramp. Furthermore,~\citeauthor{Li_hyperpol_2013} suggest a total simulation time of three to four optical cycles after the ramp and 
that field strenghts in the range $\mathcal{E}_0 \in [0.0005, 0.005]\,\text{a.u.}$ are used.

\section{Results and Discussion}
In order to assess optical properties extracted from the real-time TDOMP2 method we compute absorption spectra, polarizabilities and first hyperpolarizabilities for the ten-electron systems \ch{Ne}, \ch{HF}, \ch{H2O}, \ch{NH3}, and \ch{CH4}. 
To the best of our knowledge, response theory has neither been derived nor implemented for the OMP2 method and, therefore,
we compare results from TDOMP2 simulations with those extracted from real-time TDCCSD and TDCC2 simulations, and with 
results from CCSD and CC2 response theory (LRCCSD/LRCC2).~\cite{Koch1990,Christiansen_cc2_1995}
\added[comment={R2.7}]{We only compute polarizabilities and hyperpolarizabilities using the TDCC2-b method,
since \citeauthor{kats2006local_cc2b}~\cite{kats2006local_cc2b} found that the effect of the fully $T_1$-transformed
Fock operator on excitation energies is negligible.}

For \ch{Ne} we use the d-aug-cc-pVDZ basis set in order to compare wih \citeauthor{Larsen_polarizabilities_1999}~\cite{Larsen_polarizabilities_1999},
while for the remaining molecules we use the aug-cc-pVDZ basis set~\cite{Dunning1989,Kendall1992,Woon1994}. Basis set specifications were downloaded from the
Basis Set Exchange~\cite{Pritchard2019} and the molecular geometries used are given in the supporting information (SI).

The real-time simulations and correlated ground-state optimizations are carried out with a locally developed code described in previous
publications~\cite{Pedersen2019,Kristiansen2020,Pedersen2021} using
Hamiltonian matrix elements and HF orbitals computed with the PySCF package~\cite{Sun2018}.
The CCSD and CC2 ground states are computed with the direct inversion in the iterative subspace (DIIS)~\cite{helgaker2014molecular} procedure,
and the OMP2 ground state with the algorithm described by \citeauthor{Bozkaya_OMP2_2011}~\cite{Bozkaya_OMP2_2011} with the diagonal approximation of the Hessian.
The convergence threshold for the residual norms is set to $10^{-10}$. Ground-state energies and non-zero permanent dipole moments for the systems considered
are given in the SI. The CCSD and CC2 linear and quadratic response calculations are performed with the Dalton quantum
chemistry package~\cite{Aidas2014,Olsen2020}.

The TDOMP2, TDCCSD, \replaced[comment={R2.7}]{TDCC2, and TDCC2-b}{and TDCC2} equations of motion are integrated using the symplectic Gauss-Legendre integrator~\cite{HairerLubichWanner_GNI,Pedersen2019}. For all cases 
the integration is performed with time step $\Delta t = 0.01\,\text{a.u.}$ using the sixth-order ($s=3$) Gauss-Legendre integrator and a convergence 
threshold of $10^{-10}$ (residual norm) for the fixed-point iterations. In all RT simulations, the ground state is taken as the initial state of the system
and we use a closed-shell spin-restricted implementation of the equations. Also the response calculations are performed in the
closed-shell spin-restricted formulation.

\subsection{Absorption spectra}
Absorption spectra are computed as described in Sec.~\ref{sec:opt_props_from_rtsims} with the electric-field impulse of Eq.~\eqref{delta_pulse}.
The field strength is $\mathcal{E}_0 = 0.001\,\text{a.u.}$, which is small enough to ensure that only transitions from the ground state to
dipole-allowed excited states occur, whilst strong enough to induce numerically significant oscillations.
The induced dipole moment is recorded at each of $100\ 000$ time steps after application of the impulse, yielding 
a spectral resolution of about $0.006\,\text{a.u.}$ ($0.163\,\text{eV}$) in the FFT of Eq.~\eqref{fourier_lorentz}.
The damping parameter is $\gamma = 0.00921\,\text{a.u.}$ ($0.251\,\text{eV}$), which implies that the full width at half maximum of the Lorentzian absorption lines
is roughly $50\%$ greater than the spectral resolution. Hence, very close-lying resonances will appear as a single broader absorption line,
possibly with ``shoulders''.

The quality of TDOMP2 absorption spectra can be assessed by comparison with the well-known and highly similar TDCC2 theory
(see SI for a validation of the TDCC2 spectra by comparison with LRCC2 spectra \added[comment=R1.2]{in the range from $0\,\text{eV}$ to $930\,\text{eV}$}),
the essential difference between the two methods being how orbital relaxation is treated. In general, LRCC2 theory provides
excellent valence excitation energies, often better than those of LRCCSD theory, for states with predominant single-excitation
character, see, for example, the benchmark study by \citeauthor{Schreiber2008}~\cite{Schreiber2008}
Preliminary and rather limited tests of excitation energies computed with NOCC theory revealed
virtually no effect of the different orbital relaxation treatments~\cite{NOCC_Pedersen_2001} and, therefore, one might expect
only minor deviations between TDOMP2 and TDCC2 absorption spectra, at least in the valence regions.
For a full comparison of the two methods,
we will not limit ourselves to selected
valence-excited states but rather compare the complete spectra up to core excitations, which are also activated by the
broad-band electric-field impulse. This implies that we also compare unphysical spectral lines above the ionization threshold, which
arise artificially from the use of an incomplete basis set that ignores the electronic continuum. Furthermore, we do not use proper
core-correlated basis sets for describing core excitations, nor do we make any attempt at properly separating the core excitations from high-lying
artificial valence excitations. Hence, no direct comparison with experimental data will be done in this work.
\added[comment={R1.1}]{We instead refer to Refs.~\citenum{Bartlett_Core_2019} and \citenum{Coriani_Core_2012}, where
experimental near-edge X-ray absorption spectra 
are compared with those computed with a range of
LRCC and EOM-CC methods and large basis sets for systems studied in this work.
Importantly, the direct comparison of TDCC2 and TDOMP2 absorption spectra will indicate the effects of fully
bivariational, time-dependent orbitals on core excitations,
where orbital relaxation is expected to play a key role---see, e.g., the discussion by
\citeauthor{Bartlett_Core_2019}~\cite{Bartlett_Core_2019} for systems also considered in the present work.}

\begin{figure}
    \centering
    \subfloat{{\includegraphics[scale=1]{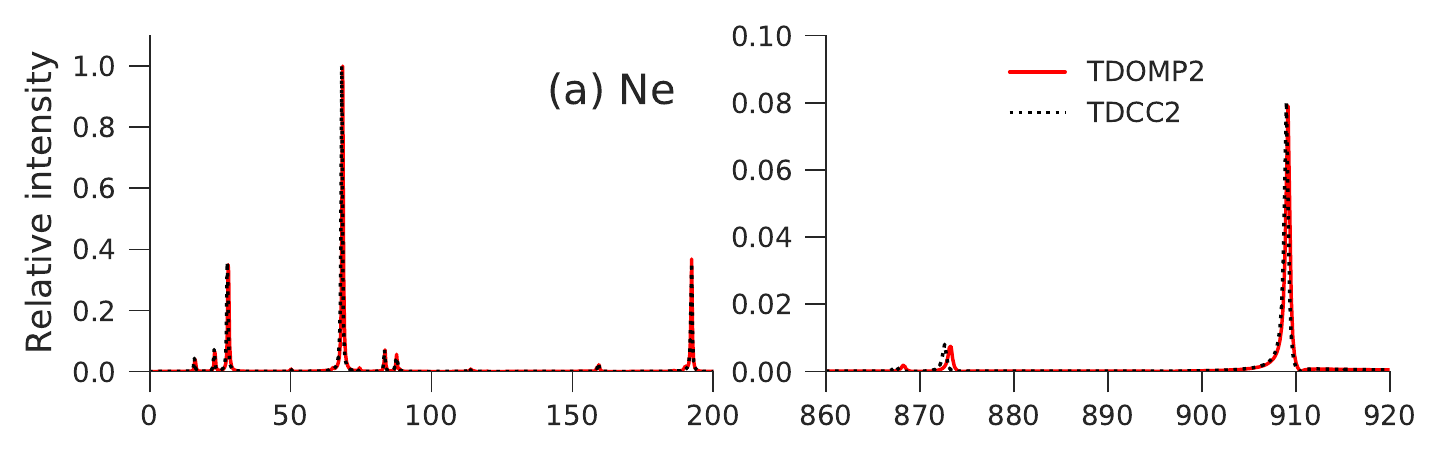} }}\\
    \vspace{-1.3\baselineskip}
    \subfloat{{\includegraphics[scale=1]{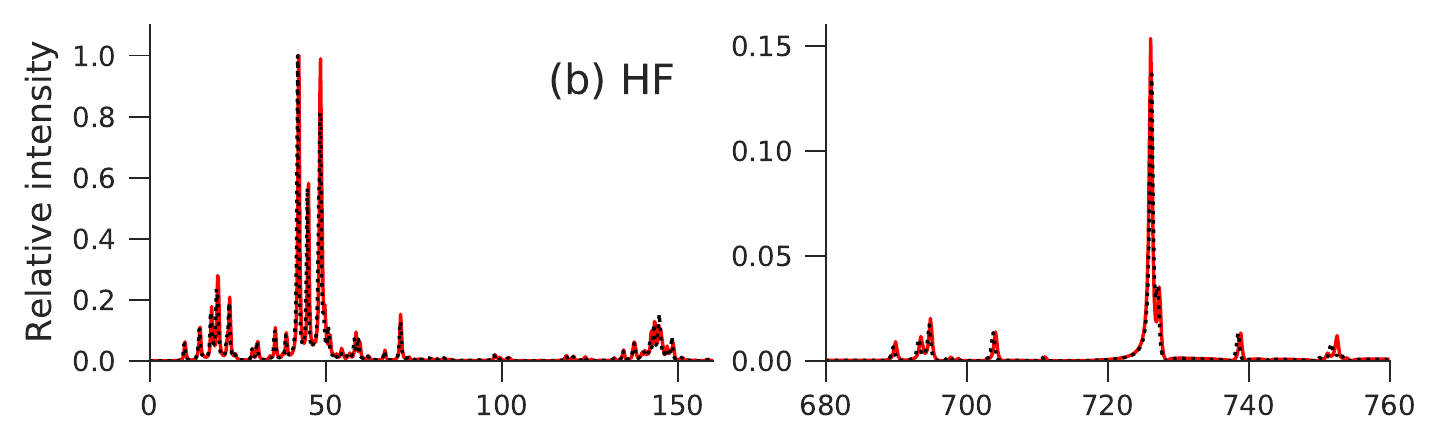} }}\\
    \vspace{-1.3\baselineskip}
    \subfloat{{\includegraphics[scale=1]{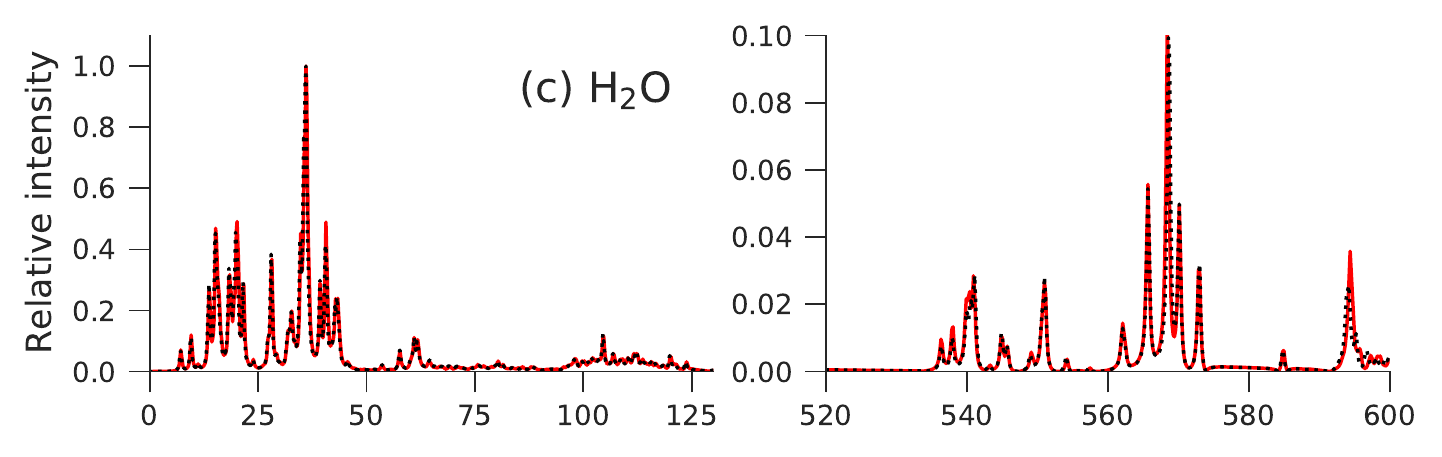} }}\\
    \vspace{-1.3\baselineskip}
    \subfloat{{\includegraphics[scale=1]{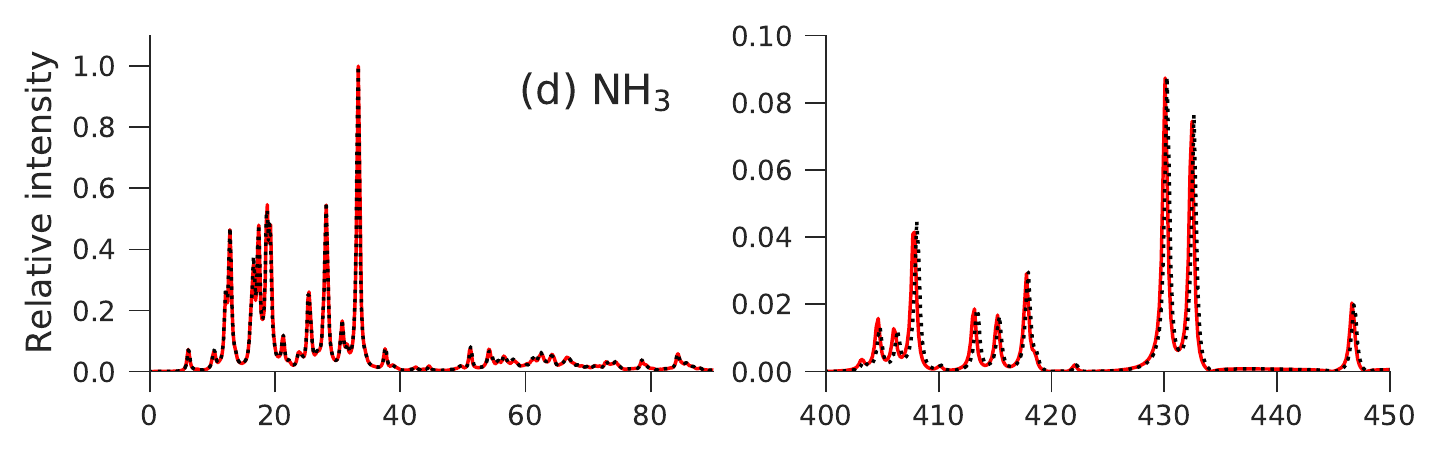} }} \\
    \vspace{-1.3\baselineskip}
    \subfloat{{\includegraphics[scale=1]{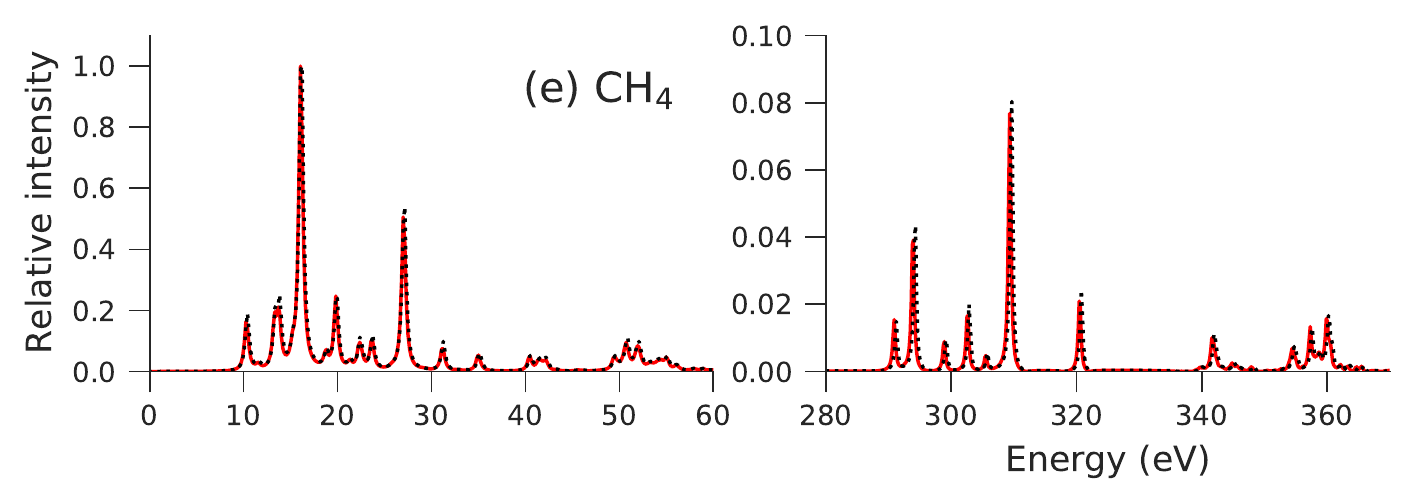} }} \\    
    \caption{Absorption spectra computed with TDOMP2 and TDCC2 for \ch{Ne}, \ch{HF}, \ch{H2O}, \ch{NH3} and \ch{CH4}.}%
    \label{fig:absorption_spectra}%
\end{figure}

In Figure~\ref{fig:absorption_spectra} we have plotted the TDOMP2 and TDCC2 electronic absorption spectra up to and including the core region.
Although deviations between the TDOMP2 and TDCC2 spectra are visible, the two methods yield very similar results both in 
the valence region and in the core region. The excitation energies identified from the simulated spectra by automated peak detection
are reported in Table~\ref{lowest_transition_freq}
for the dipole-allowed states below $30\,\text{eV}$ and confirm the close agreement of TDOMP2 and TDCC2 theory.
\begin{table}
\centering
\caption{Dipole-allowed excitation energies (in eV) below $30\,\text{eV}$ extracted from TDOMP2 and TDCC2 simulations.}
\begin{tabular}{l r r l r r l r r}
\hline 
\hline
        & TDOMP2  & TDCC2    &          & TDOMP2  & TDCC2   &         & TDOMP2  & TDCC2   \\
\hline
\ch{H2O} & $7.17$  & $7.17$  & \ch{NH3} & $6.15$  & $6.15$  & \ch{CH4}& $10.25$ & $10.42$ \\
         & $9.56$  & $9.56$  &          & $7.52$  & $7.69$  &         & $11.61$ & $11.61$ \\
         & $11.10$ & $11.10$ &          & $10.25$ & $10.25$ &         & $13.32$ & $13.49$ \\
         & $13.66$ & $13.66$ &          & $12.13$ & $12.13$ &         & $13.66$ & $13.83$ \\
         & $15.20$ & $15.20$ &          & $12.81$ & $12.81$ &         & $16.06$ & $16.23$ \\
         & $18.45$ & $18.28$ &          & $16.57$ & $16.57$ &         & $18.79$ & $18.79$ \\
         & $20.15$ & $19.81$ &          & $17.42$ & $17.42$ &         & $19.81$ & $19.81$ \\
         & $21.69$ & $21.52$ &          & $18.79$ & $18.79$ &         & $21.35$ & $21.35$ \\
         & $23.91$ & $23.74$ &          & $19.30$ & $19.13$ &         & $22.38$ & $22.38$ \\
         & $27.33$ & $27.33$ &          & $21.35$ & $21.35$ &         & $23.57$ & $23.74$ \\
         & $28.18$ & $28.01$ &          & $22.20$ & $22.20$ &         & $26.99$ & $27.16$ \\
\ch{Ne}  & $16.06$ & $15.88$ &          & $23.91$ & $23.91$ & \ch{HF} & $10.08$ & $9.91$  \\
         & $23.06$ & $22.89$ &          & $25.45$ & $25.28$ &         & $14.35$ & $14.18$ \\
         & $27.84$ & $27.50$ &          & $26.82$ & $26.82$ &         & $19.30$ & $18.96$ \\
         &         &         &          & $28.18$ & $28.18$ &         & $22.72$ & $22.54$ \\
         &         &         &          & $29.21$ & $29.21$ &         & $24.25$ & $24.08$ \\
		 &         &         &          &         &         &         & $29.21$ & $29.04$ \\
\hline
\hline
\label{lowest_transition_freq}
\end{tabular}
\end{table}
The greatest deviations are found for the HF molecule, especially for the
intensities. Some intensity deviations are expected, as the TDOMP2 method is gauge invariant (in the complete basis set limit)
while TDCC2 theory is not~\cite{Pedersen_OCC_1999,NOCC_Pedersen_2001}, which is bound to influence transition moments but not
necessarily excitation energies.
In the core regions, we note that the spectra of \ch{H2O}, \ch{NH3}, and \ch{CH4} agree qualitatively
with the core spectra obtained by \citeauthor{Bartlett_Core_2019}~\cite{Bartlett_Core_2019} from TD-EOM-CCSD theory.
\added[comment={R1.1}]{Keeping in mind that the large deviations between LRCC2/LRCCSD and experimental core excitation
energies are ascribed to missing orbital-relaxation effects, it is intriguing to observe that
the fully bivariational orbital evolution included in TDOMP2 theory hardly affects the core spectra relative to
TDCC2 theory. Using automated peak detection, we find that the differences in excitation energies in the core region between
the TDOMP2 and TDCC2 spectra are within $1$--$2$ times the spectral resolution. Since the error of LRCC2 core excitation
energies typically is several eV, we conclude that the orbital relaxation provided by TDOMP2 theory is not sufficient
to significantly improve the agreement with experimental results. This observation calls for further investigations
with larger basis sets, higher resolution (longer simulation times),
and full inclusion of double excitations (the TDOCCD and TDNOCCD methods).}

\subsection{Polarizabilities and first hyperpolarizabilities}
Polarizabilities and first hyperpolarizabilities are computed using an electric field given by Eq.~(\ref{ramp_field}).
After the initial one-cycle ramp we propagate for three optical cycles. 
The first- and second-order time-dependent dipole response functions are computed by finite difference according to Eqs.~(\ref{first_order_dip_response})
and (\ref{second_order_dip_response}), with the first optical cycle of the time evolution discarded because of the ramping.
We then perform least-squares fitting~\cite{hastie_linear_2009} of the time-domain dipole response functions to the form of Eqs.~(\ref{polarizability})
and (\ref{first-polarizability}), obtaining frequency-dependent polarizabilities
and hyperpolarizabilities.
For all systems we use the field strengths $\mathcal{E}_0 = \pm 0.0001, \pm 0.0002\,\text{a.u.}$ to compute the dipole derivatives using
finite difference.

The diagonal elements of the frequency-dependent polarizability tensor
extracted from TDCCSD, TDOMP2, \replaced[comment={R2.7}]{TDCC2, and TDCC2-b}{and TDCC2}
simulations for \ch{Ne}, \ch{HF}, \ch{H2O}, \ch{NH3}, and \ch{CH4}
are listed in Table~\ref{All:polarizabilities} along with results from LRCCSD and LRCC2 theory.
\begin{table}
\centering
\caption{Polarizabilities (a.u.) of \ch{Ne}, \ch{HF}, \ch{H2O}, \ch{NH3}, and \ch{CH4} extracted from TDCCSD, TDOMP2, TDCC2, and TDCC2-b simulations. The LRCCSD and LRCC2 results for \ch{Ne} and \ch{HF} are from Ref.~\citenum{Larsen_polarizabilities_1999} and the remaining LRCCSD and LRCC2 results are computed with the Dalton quantum chemistry program (Ref.~\citenum{Aidas2014}).}
\begin{tabular}{l l r r r r r r r r r}
\hline 
\hline
\ch{Ne}&$\omega\,(\text{a.u.})$ & $0.1$ & $0.2$ & $0.3$ & $0.4$ & $0.5$ \\
\hline
&LRCCSD   & $2.74$ & $2.83$ & $3.01$ & $3.38$ & $4.23$\\		 
&TDCCSD   & $2.74$ & $2.83$ & $3.03$ & $3.49$ & $4.76$\\
&TDOMP2   & $2.77$ & $2.87$ & $3.07$ & $3.58$ & $4.99$\\
&LRCC2    & $2.86$ & $2.96$ & $3.18$ & $3.59$ & $4.74$\\
&TDCC2    & $2.87$ & $2.98$ & $3.19$ & $3.75$ & $5.29$\\
&TDCC2-b   & $2.86$ & $2.97$ & $3.18$ & $3.73$ & $5.26$\\
\hline
\ch{HF}&$\omega\,(\text{a.u.})$        & \multicolumn{2}{c}{$0.1$}&  \multicolumn{2}{c}{$0.2$} & \multicolumn{2}{c}{$0.3$}            \\ 
&                      & $\alpha_{yy}$   & $\alpha_{zz}$ & $\alpha_{yy}$   & $\alpha_{zz}$ & $\alpha_{yy}$   & $\alpha_{zz}$ \\
\hline 
&LRCCSD  &     $4.44$      &   $6.41$  &     $4.83$      &  $6.83$ &     $6.19$      &   $7.73$          \\
&TDCCSD   & $4.45$ & $6.41$ & $4.84$ & $6.83$ & $6.72$ & $7.84$\\
&TDOMP2   & $4.56$ & $6.49$ & $5.03$ & $6.94$ & $7.71$ & $7.96$\\
&LRCC2   &     $4.70$      &	$6.78$  &     $5.20$     &  $7.25$ &     $7.24$      &   $8.29$     	 	\\
&TDCC2    & $4.75$ & $6.85$ & $5.28$ & $7.36$ & $8.54$ & $8.45$\\
&TDCC2-b   & $4.72$       &   $6.79$  & $5.24$       &  $7.28$      &    $8.42$ & $8.36$   \\
\hline
\ch{H2O} & $\omega\,(\text{a.u.})$ & \multicolumn{3}{c}{$0.0428$} & \multicolumn{3}{c}{$0.0656$} & \multicolumn{3}{c}{$0.1$} \\
&                      & $\alpha_{xx}$          & $\alpha_{yy}$   & $\alpha_{zz}$ & $\alpha_{xx}$  & $\alpha_{yy}$   & $\alpha_{zz}$ & $\alpha_{xx}$  & $\alpha_{yy}$   & $\alpha_{zz}$ \\
\hline 
&LRCCSD  &     $8.78$    &     $9.93$       &   $9.11$      &     $8.89$    &     $9.99$       &   $9.19$      &     $9.18$    &     $10.14$      &   $9.37$\\
&TDCCSD   & $8.78$ & $9.93$ & $9.11$ & $8.90$ & $10.00$ & $9.19$ & $9.19$ & $10.14$ & $9.37$\\
&TDOMP2   & $9.16$ & $10.06$ & $9.34$ & $9.29$ & $10.13$ & $9.42$ & $9.62$ & $10.27$ & $9.63$\\
&LRCC2   &     $9.41$    &     $10.43$      &   $9.63$      &     $9.55$    &     $10.50$      &   $9.71$      &     $9.91$    &     $10.66$      &   $9.92$\\		 	
&TDCC2    & $9.51$ & $10.56$ & $9.74$ & $9.65$ & $10.63$ & $9.83$ & $10.01$ & $10.79$ & $10.06$\\
&TDCC2-b   & $9.44$       &  $10.47$      &   $9.66$     &   $9.58$     &  $10.54$ & $9.74$  & $9.94$ & $10.71$ & $9.97$  \\
\hline
\ch{NH3} & $\omega\,(\text{a.u.})$        & \multicolumn{2}{c}{$0.0428$} & \multicolumn{2}{c}{$0.0656$} & \multicolumn{2}{c}{$0.1$} \\
&        & $\alpha_{yy}$   & $\alpha_{zz}$ & $\alpha_{yy}$& $\alpha_{zz}$ & $\alpha_{yy}$& $\alpha_{zz}$ \\
\hline 
&LRCCSD  &      $13.10$    &    $15.04$    & $13.20$      &   $15.35$     & $13.44$      & $16.15$       \\
&TDCCSD   & $13.10$  & $15.05$ & $13.20$ & $15.36$ & $13.45$ & $16.15$\\
&TDOMP2   & $13.23$  & $15.60$ & $13.34$ & $15.98$ & $13.59$ & $16.95$\\
&LRCC2   &      $13.56$    &    $15.86$    & $13.67$      &   $16.21$     & $13.92$      & $17.15$       \\
&TDCC2    & $13.72$  & $16.03$ & $13.83$ & $16.43$ & $14.10$ & $17.40$\\
&TDCC2-b   & $13.64$  & $15.93$ &  $13.75$      &  $16.32$      &  $14.01$ & $17.28$    \\
\hline
\ch{CH4}&$\omega\,(\text{a.u.})$ & $0.0656$ & $0.1$ & $0.2$\\
\hline
&LRCCSD  & $17.05$ & $17.39$ & $19.55$ \\ 
&TDCCSD   & $17.05$ & $17.39$ & $19.58$\\
&TDOMP2   & $17.18$ & $17.53$ & $19.79$\\
&LRCC2   & $17.49$ & $17.84$ & $20.08$ \\
&TDCC2    & $17.69$ & $18.05$ & $20.34$\\
&TDCC2-b   & $17.61$ & $17.96$ &   $20.25$\\
\hline 
\hline
\end{tabular}
\label{All:polarizabilities}
\end{table}
All three diagonal elements are identical by symmetry for \ch{Ne} and \ch{CH4}, $\alpha_{xx} = \alpha_{yy}$ for \ch{HF} and \ch{NH3},
and off-diagonal elements vanish for all systems considered here.
The polarizability diverges at the (dipole-allowed) excitation energies and, therefore, we select frequencies below the first
dipole-allowed transition in Table~\ref{lowest_transition_freq} (roughly $0.6\,\text{a.u.}$ for \ch{Ne},
$0.4\,\text{a.u.}$ for \ch{HF}, $0.3\,\text{a.u.}$ for \ch{H2O}, $0.2\,\text{a.u.}$ for \ch{NH3}, and $0.4\,\text{a.u.}$ for \ch{CH4}). 

The benchmark study by \citeauthor{Larsen_polarizabilities_1999}~\cite{Larsen_polarizabilities_1999}
indicated that LRCCSD theory yields accurate static and dynamic polarizabilities, although triple excitations are
needed to obtain results very close to FCI theory, whereas results from LRCC2 theory are significantly
less accurate. Our results in Table~\ref{All:polarizabilities} confirm this finding in the sense that TDCC2 (and LRCC2) results are 
quite far from the corresponding TDCCSD (and LRCCSD) results. We also note that TDCCSD and LRCCSD results agree to a much greater
extent than the results from TDCC2 and LRCC2 theory.

Unfortunately, we have not been able to identify the source of this behavior of the TDCC2 model.
The agreement between the results from simulations and from response theory generally worsen as the frequency approaches
the lowest-lying dipole-allowed transition. In this ``semi-transparent'' regime, the assumptions of linear response theory
are violated and the first-order time-dependent induced dipole moment can not be described as the simple function in
Eq.~(\ref{polarizability}).
\begin{figure}
    \centering
    \subfloat{{\includegraphics[scale=1]{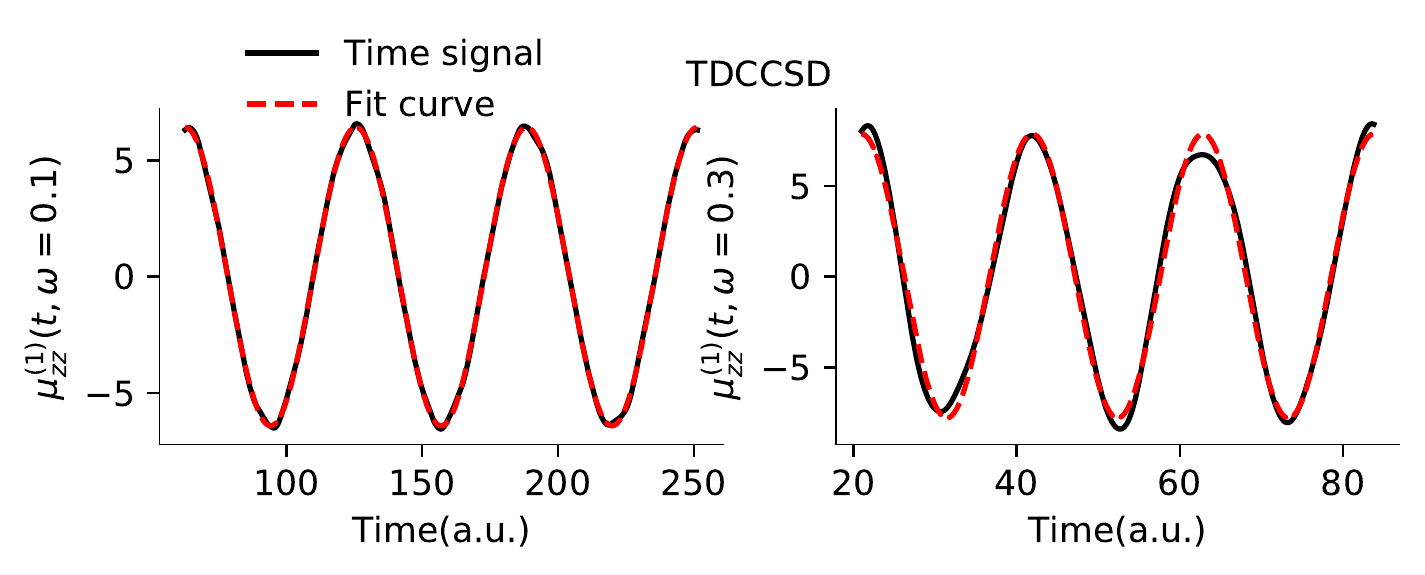} }}\\
    \vspace{-1.3\baselineskip}
    \subfloat{{\includegraphics[scale=1]{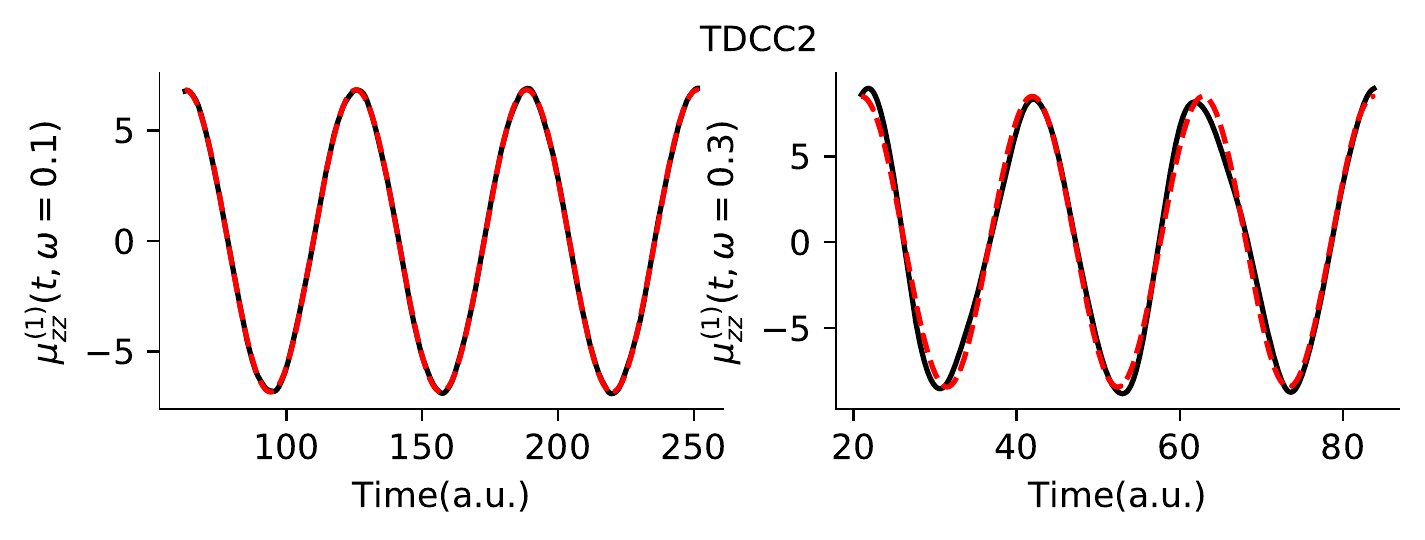} }}\\
    \vspace{-1.3\baselineskip}
    \subfloat{{\includegraphics[scale=1]{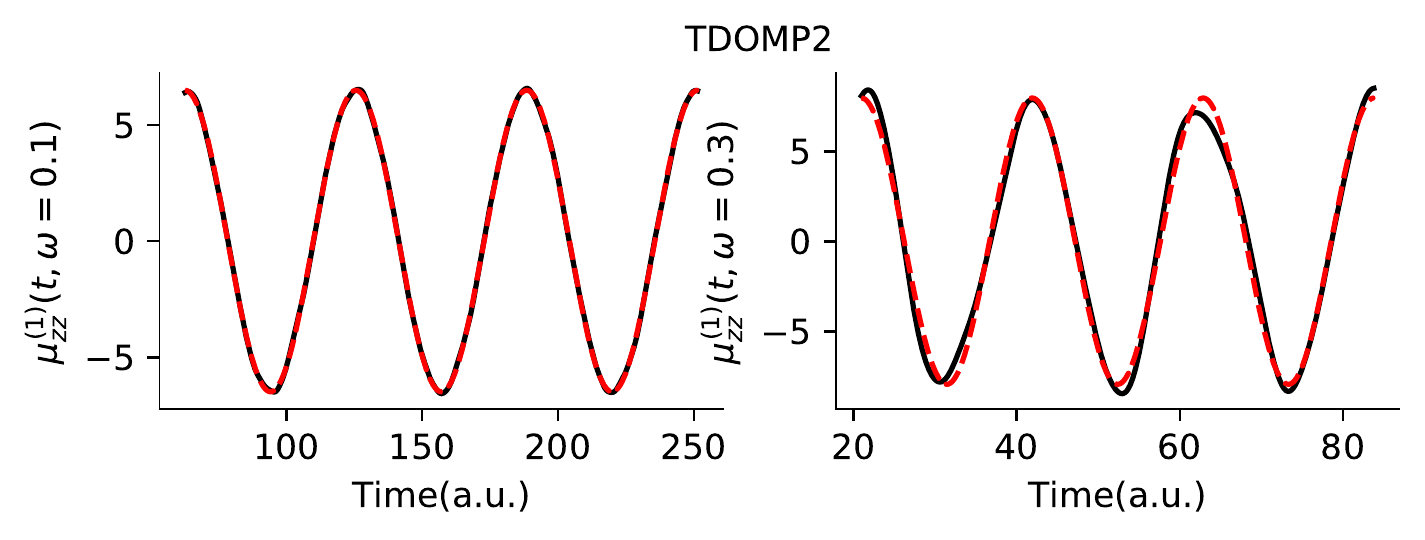} }}\\
    \caption{\added{The $\text{zz}$-component} of the first-order dipole responses for \ch{HF} at $\omega = 0.1\,\text{a.u.}$ and $\omega=0.3\,\text{a.u.}$ from TDCCSD, TDCC2, and TDOMP2 simulations.}%
    \label{fig:first_order_dip_responses}%
\end{figure}

\begin{figure}
    \centering
    \subfloat{{\includegraphics[scale=1]{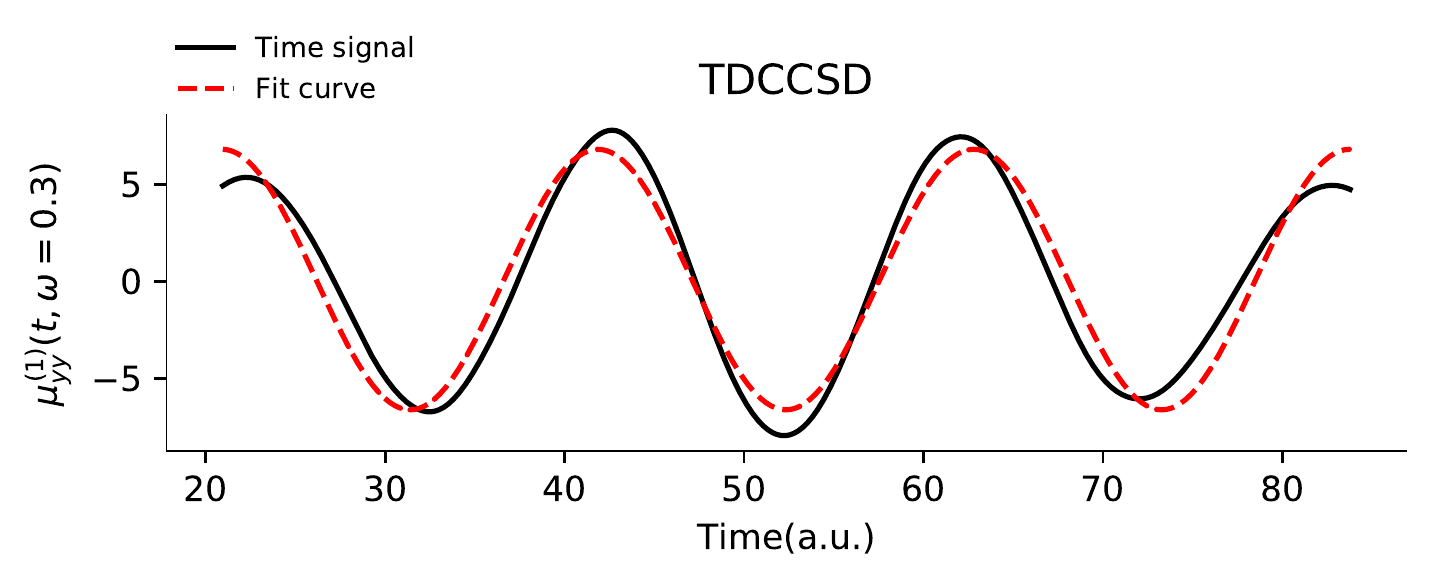} }}\\
    \vspace{-1.3\baselineskip}
    \subfloat{{\includegraphics[scale=1]{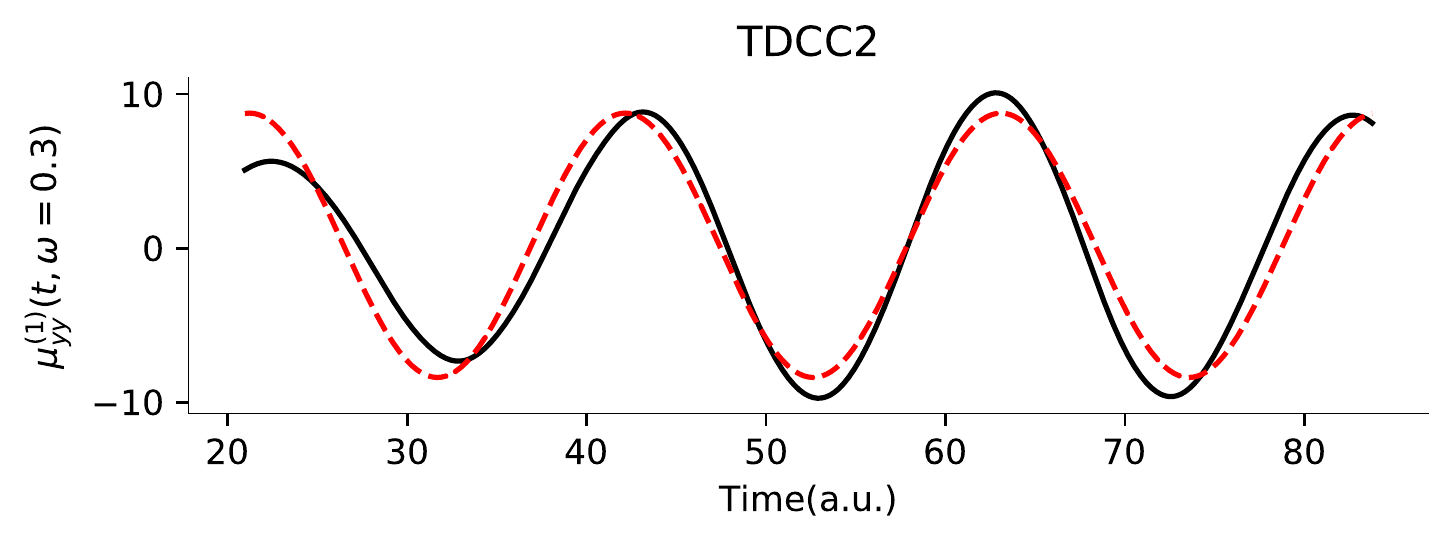} }}\\
    \vspace{-1.3\baselineskip}
    \subfloat{{\includegraphics[scale=1]{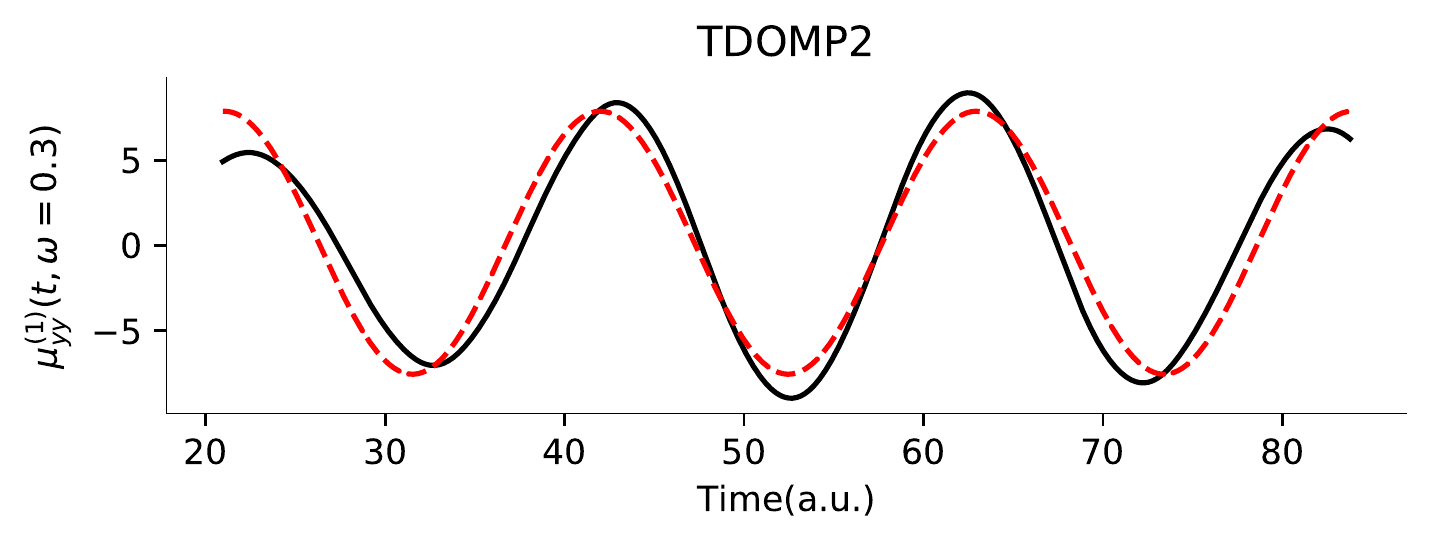} }}\\
    \caption{\added{The $\text{yy}$-component of the first-order dipole responses for \ch{HF} at $\omega=0.3\,\text{a.u.}$ from TDCCSD, TDCC2, and TDOMP2 simulations.}}%
    \label{fig:first_order_dip_responses_yy_hf}%
\end{figure}

This is confirmed by the plots of simulated time signals and the least-squares fits in Fig.~\ref{fig:first_order_dip_responses}
where the former clearly can only be accurately described by Eq.~(\ref{polarizability}) at sufficiently low (transparent) frequencies.
The TDCC2 least-squares fits, however, do not appear worse than those of TDCCSD or TDOMP2 theory. Hence, larger deviations from the
form in Eq.~\eqref{polarizability} can not explain the discrepancies between TDCC2 and LRCC2 results.

\added[comment={R1.3}]{Furthermore, we note the relatively large discrepancy between the LRCCSD and TDCCSD results for the \ch{HF} molecule at $\omega=0.3$ a.u. and the \ch{Ne} atom at $\omega=0.4$ a.u. and $\omega=0.5$ a.u. 
In these cases the first-order response function extracted from the time-dependent simulations~(\ref{first_order_dip_response}) for all methods considered does not agree with the assumption of a pure cosine wave~(\ref{polarizability}), as shown in Fig.~\ref{fig:first_order_dip_responses_yy_hf} for the \ch{HF} molecule. The source of deviation is a combined effect of proximity to a pole, non-adiabatic effects arising from ramping up the field over a single cycle, and the absence of higher-order corrections in the finite-difference expressions for the response functions~\cite{Li_hyperpol_2013}. This is also likely to be the source of the irregular behavior of $\alpha_{yy}$ computed with the TDCC2 and TDCC2-b methods. 
}

\begin{figure}
    \centering
    \includegraphics{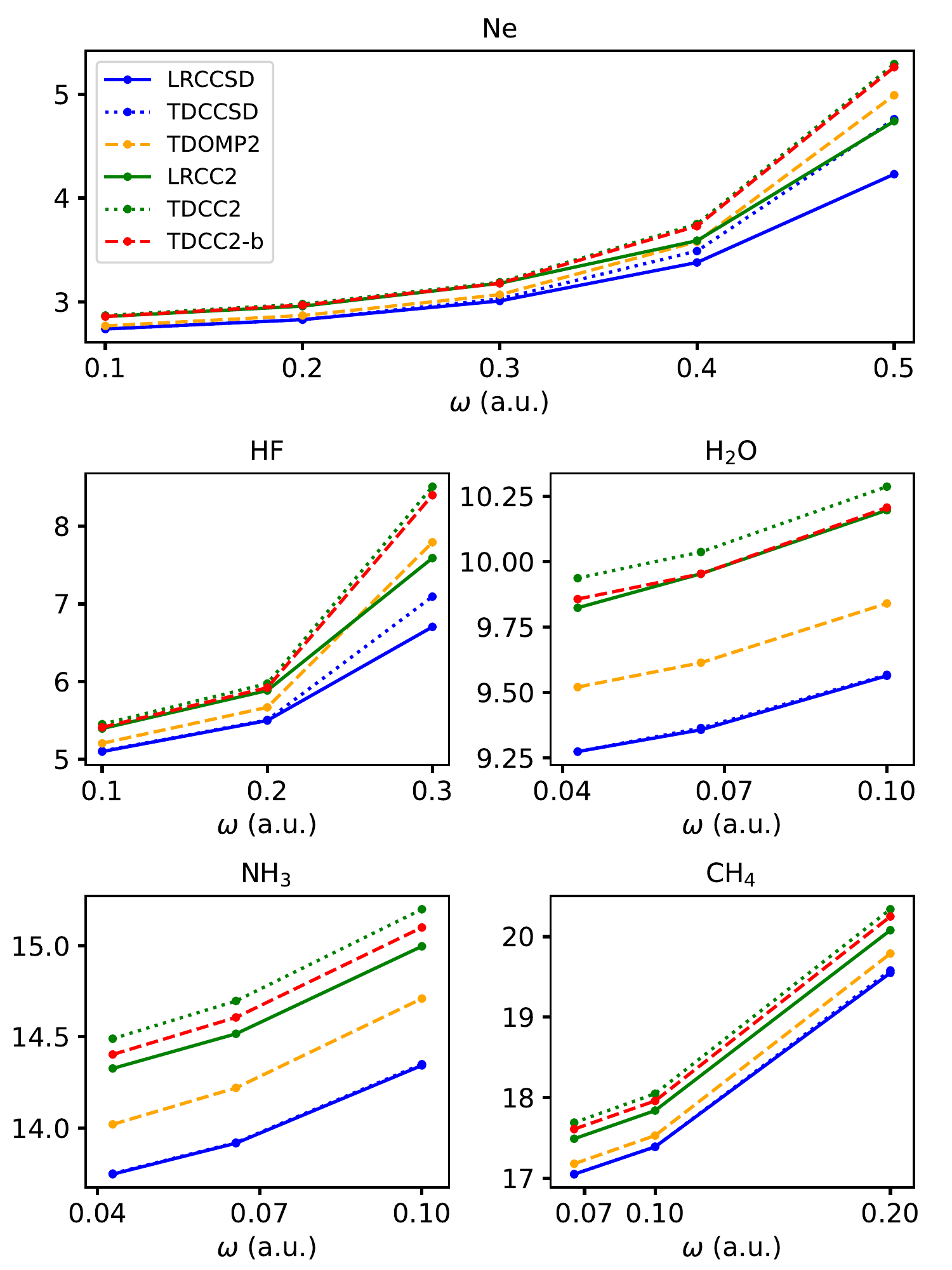}
    \caption{Isotropic polarizabilities extracted from TDCC2, TDCC2-b, TDOMP2, and TDCCSD simulations, and from
             LRCC2 and LRCCSD calculations.}
    \label{fig:isopol}
\end{figure}

Interestingly, we observe that polarizabilities from TDOMP2 theory are generally in better agreement with the TDCCSD values than 
those from TDCC2 (and LRCC2) theory. This trend is particularly evident from Fig.~\ref{fig:isopol} where we have plotted the
dispersion of the isotropic polarizability, $\alpha_\text{iso} = (\alpha_{xx} + \alpha_{yy} + \alpha_{zz})/3$.
Keeping in mind the similarity between the TDOMP2 and TDCC2 spectra,
the pronounced difference between TDOMP2 and TDCC2 polarizabilities is somewhat surprising.
It is, however, in agreement with the observation by \citeauthor{Larsen_polarizabilities_1999}~\cite{Larsen_polarizabilities_1999} that
orbital relaxation has a sizeable impact on polarizabilities within CC theory, albeit not always improving the results relative to FCI calculations.
Only static polarizabilities were considered by
\citeauthor{Larsen_polarizabilities_1999}~\cite{Larsen_polarizabilities_1999} since the orbital relaxation---formulated as a variational
HF constraint within conventional CC response theory---leads to spurious uncorrelated poles in the response functions, making it useless for
dynamic polarizabilities. The orbitals are treated as fully bivariational variables within TDOMP2 theory and, consequently,
spurious poles are avoided.~\cite{Pedersen_OCC_1999}
Our results, therefore, seem to indicate that a fully bivariational treatment of orbital relaxation is beneficial for polarizability predictions.

\added[comment={R2.7}]{The partial orbital relaxation included in the TDCC2-b method does not yield equally good polarizabilities.
In most cases, the results are nearly identical to the TDCC2 ones, except for the \ch{H2O} and \ch{NH3} molecules where the TCCC2-b 
polarizabilities are closer to the LRCC2 results, see Fig.~\ref{fig:isopol}.}

In Table~\ref{All:first_hyper_polarizabilities} we list frequency-dependent first hyperpolarizabilities for \ch{HF}, \ch{H2O}, and \ch{NH3}.
Only the nonvanishing diagonal components of the practically most important response tensors corresponding to optical rectification (OR), $\beta^\text{OR}_{iii} =
\beta_{iii}(0,\omega,-\omega)$,
and second harmonic generation (SHG), $\beta^\text{SHG}_{iii} = \beta_{iii}(-2\omega,\omega,\omega)$, are computed.
Formally expressable as a double summation over all excited states, the first hyperpolarizability generally requires
a high-level description of electron correlation effects for accurate calculations~\cite{Christiansen2006}.
This is reflected in our results by the relatively large difference between the TDCC2 and TDCCSD methods.
\begin{table}
\centering
\caption{First hyperpolarizabilities (a.u.) of \ch{HF}, \ch{H2O}, and \ch{NH3} from TDCCSD, TDOMP2, TDCC2, and TDCC2-b simulations.
Notation: $\beta_{iii}^\text{OR} = \beta_{iii}(0;\omega,-\omega)$ and 
$\beta_{iii}^\text{SHG} = \beta_{iii}(-2\omega;\omega,\omega)$. The LRCCSD and LRCC2 results for \ch{HF} are taken from
\citeauthor{Larsen_polarizabilities_1999}~\cite{Larsen_polarizabilities_1999}
}
\begin{tabular}{l l r r r r r r r r}
\hline 
\hline
\ch{HF} &$\omega\,(\text{a.u.})$        & \multicolumn{2}{c}{$0.1$} & \multicolumn{2}{c}{$0.2$} & \multicolumn{2}{c}{$0.3$} \\
&                      & $\beta_{zzz}^\text{OR}$ & $\beta_{zzz}^\text{SHG}$ & $\beta_{zzz}^\text{OR}$ & $\beta_{zzz}^\text{SHG}$ & $\beta_{zzz}^\text{OR}$ & $\beta_{zzz}^\text{SHG}$ \\
\hline 
&LRCCSD  & $12.81$           & $14.38$ & $15.28$           & $29.40$ & $21.86$           & $-229.70$  \\
&TDCCSD   & $12.89$ & $14.45$ & $15.63$ & $29.32$ & $25.11$ & $-73.94$\\
&TDOMP2   & $13.05$ & $14.66$ & $15.21$ & $28.16$ & $24.98$ & $-65.73$\\
&LRCC2   & $15.52$           & $17.52$ & $18.69$           & $37.67$ & $27.35$           & $-51.78$\\
&TDCC2    & $16.53$ & $18.63$ & $19.40$ & $36.39$ & $32.11$ & $-61.17$\\
&TDCC2-b   & $15.32$        &  $17.26$       &    $17.95$     &   $33.56$      &   $29.76$      &  $-64.95$       \\ 	
\hline
\ch{H2O}&$\omega\,(\text{a.u.})$        & \multicolumn{2}{c}{$0.0428$} & \multicolumn{2}{c}{$0.0656$} & \multicolumn{2}{c}{$0.1$} \\
&                      & $\beta_{zzz}^\text{OR}$    & $\beta_{zzz}^\text{SHG}$ & $\beta_{zzz}^\text{OR}$    & $\beta_{zzz}^\text{SHG}$ & $\beta_{zzz}^\text{OR}$ & $\beta_{zzz}^\text{SHG}$ \\
\hline 
&LRCCSD                & $-9.11$   & $-9.59$ & $-9.43$   & $-10.72$ & $-10.25$  & $-14.52$\\
&TDCCSD   & $-9.14$ & $-9.62$ & $-9.50$ & $-10.78$ & $-10.47$ & $-14.69$\\
&TDOMP2   & $-9.92$ & $-10.49$ & $-10.33$ & $-11.80$ & $-11.57$ & $-17.63$\\
&LRCC2                 & $-12.39$  & $-13.12$& $-12.87$  & $-14.83$ & $-14.11$  & $-20.76$\\		 	
&TDCC2    & $-13.63$ & $-14.42$ & $-14.17$ & $-16.18$ & $-15.75$ & $-23.70$\\
&TDCC2-b   & $-11.89$ & $-12.58$ & $-12.38$ & $-14.15$ & $-13.84$ & $-21.01$\\
\hline
\ch{NH3}&$\omega\,(\text{a.u.})$ & \multicolumn{4}{c}{$0.0428$} & \multicolumn{4}{c}{$0.0656$} \\
        &               & $\beta_{yyy}^\text{OR}$ & $\beta_{yyy}^\text{SHG}$ & $\beta_{zzz}^\text{OR}$ & $\beta_{zzz}^\text{SHG}$ & $\beta_{yyy}^\text{OR}$ & $\beta_{yyy}^\text{SHG}$ & $\beta_{zzz}^\text{OR}$ & $\beta_{zzz}^\text{SHG}$ \\
\hline 
&LRCCSD  & $-14.90$ & $-15.50$ & $23.90$ & $28.02$ & $-15.30$ & $-16.88$ & $26.57$ & $40.49$ \\
&TDCCSD   & $-14.94$ & $-15.59$ & $24.20$ & $28.45$ & $-15.47$ & $-17.27$ & $27.35$ & $41.94$\\
&TDOMP2   & $-15.64$ & $-16.42$ & $30.66$ & $36.26$ & $-15.81$ & $-17.39$ & $35.50$ & $58.38$\\
&LRCC2   & $-16.69$ & $-17.40$ & $33.80$ & $39.87$ & $-17.16$ & $-19.01$ & $37.72$ & $58.61$ \\
&TDCC2    & $-17.32$ & $-18.13$ & $35.80$ & $41.90$ & $-17.51$ & $-19.17$ & $41.24$ & $66.61$\\
&TDCC2-b   & $-17.00$ & $-17.79$ & $32.60$ & $38.26$ & $-17.19$ & $-18.81$ & $37.80$ & $61.67$\\	
\hline 
\hline
\end{tabular}
\label{All:first_hyper_polarizabilities}
\end{table}
While $\beta^\text{OR}_{iii}$ is singular when the magnitude of the radiation frequency $\omega$ equals an excitation energy of the molecule,
$\beta^\text{SHG}_{iii}$ has an additional set of poles at half the excitation energies. The $\beta^\text{SHG}_{zzz}$ results at $\omega=0.3\,\text{a.u.}$
for the \ch{HF} molecule in Table~\ref{All:first_hyper_polarizabilities} are past the first pole and, hence, the sign has changed compared with
the SHG results at lower frequencies.
\added[comment={R1.4}]{The large negative value of $\beta^\text{SHG}_{zzz}$ at $\omega=0.3\,\text{a.u.}$
obtained with the LRCCSD method for the HF molecule
is due to proximity to two dipole-allowed,
$z$-polarized excitations at $0.598\,\text{a.u.}$ (oscillator strength $0.005$) and at $0.532\,\text{a.u.}$
(oscillator strength $0.157$).}

\begin{figure}
    \centering
    \subfloat{{\includegraphics[scale=1]{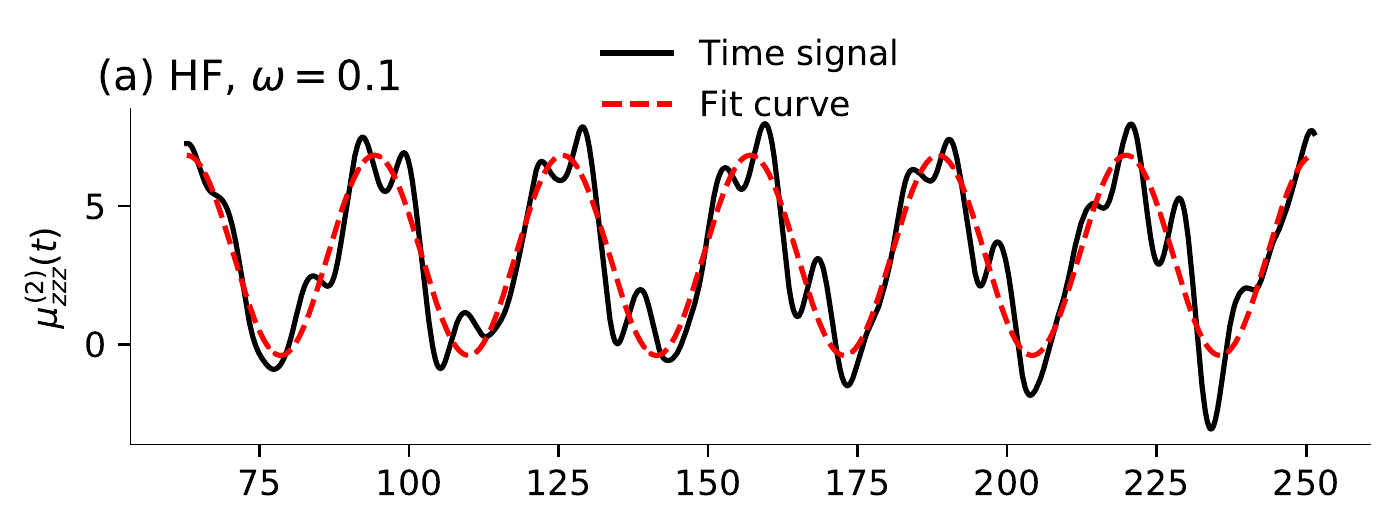} }}\\
    \vspace{-1.3\baselineskip}
    \subfloat{{\includegraphics[scale=1]{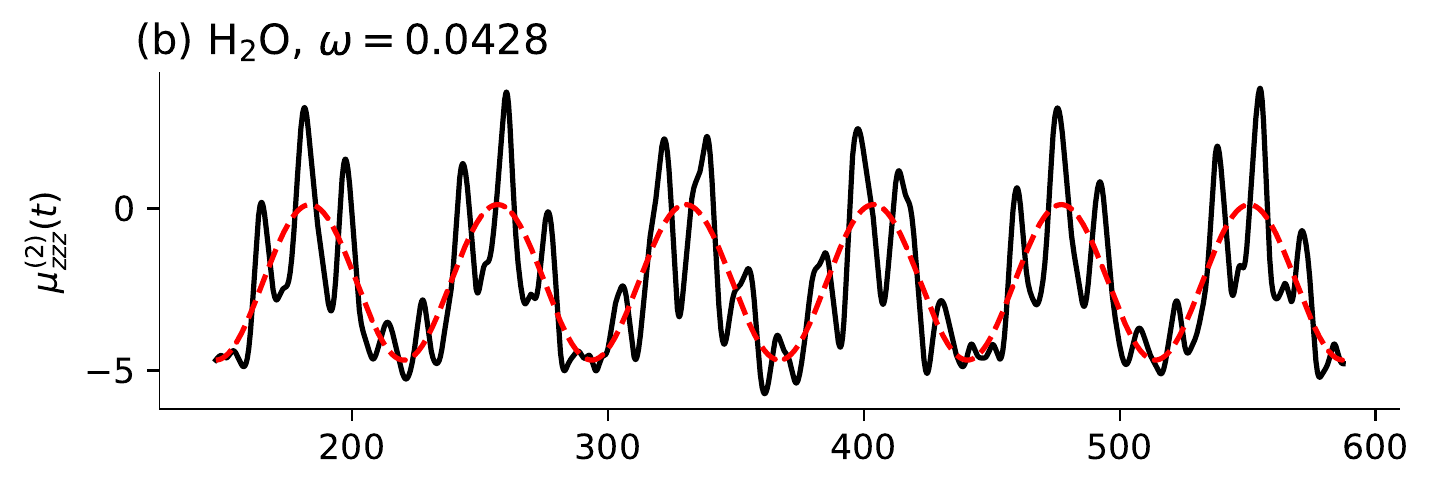} }}\\
    \vspace{-1.3\baselineskip}
    \subfloat{{\includegraphics[scale=1]{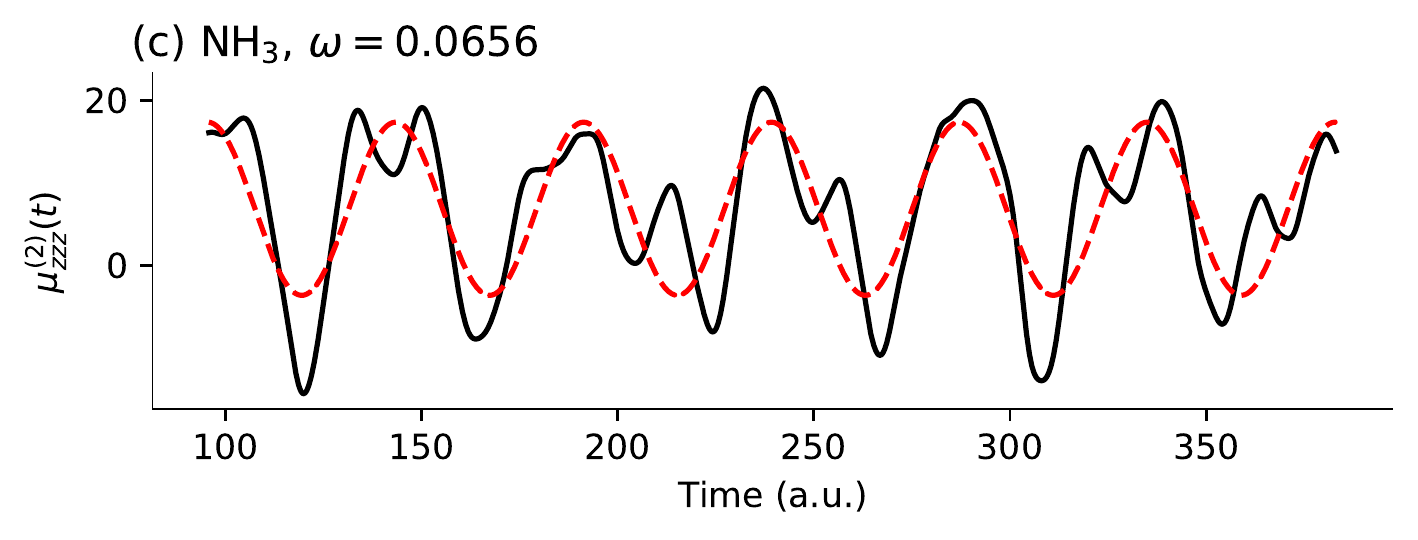} }}\\
    \caption{Second-order dipole responses for \ch{HF}, \ch{H2O}, and \ch{NH3} from TDCCSD simulations.}
	\label{fig:second_order_dip_responses}
\end{figure}

The agreement between real-time simulations and response theory is seen to be somewhat worse than for polarizabilities, especially for frequencies closer to
a pole of the hyperpolarizability. This can to a large extent be ascribed to the
second-order dipole response extracted from the time-dependent simulations not being well described by the sinusoidal form of
Eq.~(\ref{first-polarizability}), as illustrated in Fig.~\ref{fig:second_order_dip_responses}.
Analogous observations were done by \citeauthor{Li_hyperpol_2013}~\cite{Li_hyperpol_2013} in the context of
real-time time-dependent density-functional theory simulations. Hence, moving on to higher-order nonlinear optical properties
can not generally be expected to provide more than a rough estimate with the present extraction algorithm.

As for the polarizabilites above, we observe that first hyperpolarizabilities obtained from TDOMP2 simulations are generally closer to
TDCCSD and LRCCSD results than those from TDCC2 and LRCC2 theory.
The source of the improvement over TDCC2 theory must be the bivariational orbital relaxation, although we stress that the larger differences
between TDOMP2 theory and TDCCSD theory, which are particularly pronounced for \ch{NH3},
clearly demonstrate the insufficient electron-correlation treatment of the former for highly accurate predictions of nonlinear optical properties.
\added[comment={R2.7}]{The importance of orbital relaxation is corroborated by the TDCC2-b hyperpolarizabilities, which are somewhat
closer to the TDOMP2 and TDCCSD results than the TDCC2 ones.}

\section{Concluding remarks}
In this work we have presented a new unified derivation of TDOCC and TDNOCC theories, including the second-order approximations TDOMP2
and TDNOMP2, using exponential orbital-rotation operators and the bivariational Euler-Lagrange equations.
Using five small 10-electron molecules as test cases, we have extracted absorption spectra and frequency-dependent polarizabilities and
hyperpolarizabilities from TDOMP2 simulations with weak fields within the electric-dipole approximation and compared
the results with those from conventional TDCCSD and TDCC2 simulations.
While the TDOMP2 absorption spectra are almost identical to TDCC2 spectra,
\added[comment={R1.1}]{including in the spectral region of core excitations,} the TDOMP2 polarizabilities and hyperpolarizabilities are significantly
closer to TDCCSD results than those from TDCC2 simulations, especially for frequencies comfortably away from resonances.
\added[comment={R2.7}]{Further corroborated by TDCC2-b simulations, our results strongly indicate that fully (bi-)variational
orbital relaxation is important for frequency-dependent polarizabilities and hyperpolarizabilities, whilst
nearly irrelevant for absorption spectra.}

Combined with the observations by \citeauthor{OMP2_Sato_I}~\cite{OMP2_Sato_I}, who found that TDOMP2 theory outperforms TDCC2 theory for
strong-field many-electron dynamics, our results may serve as a motivation for further development of TDOMP2 theory. First of all,
a reduced-scaling implementation of TDOMP2 theory, obtained, for example, by exploiting sparsity of the correlating doubles amplitudes,~\cite{Crawford2019} 
can provide reasonably accurate results for larger systems and basis sets that are out of reach for today's TDCC implementations.
Second, an efficient implementation of OMP2 linear and quadratic response functions is warranted.

\begin{acknowledgement}
This work was supported by the Research Council of Norway through its Centres of Excellence scheme, project number 262695,
by the European Research Council under the European Union Seventh
Framework Program through the Starting Grant BIVAQUM, ERC-STG-2014 grant
agreement No. 639508, and by
the Norwegian Supercomputing Program (NOTUR) through a grant of computer time (Grant No.\ NN4654K).
\added{SK and TBP acknowledge the support of the Centre for Advanced Study in Oslo, Norway, which funded and hosted our CAS research project 
\emph{Attosecond Quantum Dynamics Beyond the Born-Oppenheimer Approximation}
during the academic year of 2021/2022.}
We thank \added[comment=R1.2]{Prof. Sonia Coriani for helpful discussions and for
providing the Lanczos-driven LRCC2 results reported in the Supporting Information, and}
Mr. Sindre Bj{\o}rings{\o}y Johnsen for making the cover art. 
\end{acknowledgement}

\begin{suppinfo}
Algebraic expressions for the closed-shell spin-restricted OMP2 method, molecular geometries, 
electronic ground-state energies and electric-dipole moments,
and comparison of TDCC2 absorption spectra with those from LRCC2 theory in the range $0$--$930\,\text{eV}$.
This information is available free of charge via the Internet at \url{https://pubs.acs.org}.
\end{suppinfo}

\bibliography{manuscript}


\end{document}



\begin{abstract}
The Supporting Information gives algebraic expressions for the closed-shell spin-restricted OMP2 method, molecular geometries,
electronic ground-state energies and dipole moments,
and a comparison of TDCC2 absorption spectra with those from LRCC2 theory from $0$--$930\,\text{eV}$.
\end{abstract}

\section{Closed-shell spin-restricted OMP2 expressions}
We now assume closed-shell systems where each orbital is doubly occupied. Then the expression for the fock matrix is given by
\begin{equation}
f^p_q = h^p_q + 2u^{pj}_{qj}-u^{pj}_{jq}.
\end{equation}

Using the following biorthogonal paramterization of $\hat{\Lambda}, \hat{T}$, 
\begin{align*}
\hat{T}_2 &= \frac{1}{2} \sum_{abij} \tau^{ab}_{ij} \hat{E}^a_i \hat{E}^b_j, \\
\hat{\Lambda}_2 &= \frac{1}{2} \sum_{abij} \lambda^{ij}_{ab}\left( \frac{1}{3} \hat{E}^j_b \hat{E}^i_a + \frac{1}{6} \hat{E}^i_b \hat{E}^j_a \right),  
\end{align*}
the derivatives of the OMP2 Hamilton function w.r.t $(\lambda^{ij}_{ab}, \kappa^i_a)$ are given by  
\begin{align}
\frac{\partial \mathcal{H}}{\partial \lambda^{ij}_{ab}} &= u^{ab}_{ij} + P^{ab}_{ij} \left( f^a_c \tau^{cb}_{ij} - f^k_i \tau^{ab}_{kj} \right),  \\
\frac{\partial \mathcal{H}}{\partial \kappa^i_a} &= h^b_i \gamma^a_b - h^a_j \gamma^j_i +  u^{pq}_{ir}\Gamma^{ar}_{pq}-u^{aq}_{rs}\Gamma^{rs}_{iq}.
\end{align}
Furthermore, one can show that 
\begin{equation}
\lambda^{ij}_{ab} = 2(2\tau^{ab}_{ij}-\tau^{ab}_{ji})^*.
\end{equation}

The expressions for the one- and two-body density matrices are given by 
\begin{align}
\gamma_i^j &= 2\delta^j_i + (\gamma_c)^j_i, \ \ (\gamma_c)^j_i = - \lambda^{kj}_{ab}\tau^{ab}_{ki} \\
\gamma^b_a &= \lambda^{ij}_{ac} \tau^{bc}_{ij}
\end{align}
and 
\begin{align}
\Gamma^{kl}_{ij} &= 4\delta^k_i\delta^l_j - 2 \delta^l_i\delta^k_j + \hat{P}^{kl}_{ij} \left(-2\delta^k_i(\gamma_{c})^l_j + \delta^k_j(\gamma_{c})^l_i \right), \\
\Gamma^{ab}_{ij} &= 2(2\tau^{ab}_{ij}-\tau^{ab}_{ji}), \\
\Gamma^{ij}_{ab} &= \lambda^{ij}_{ab} = 2(2\tau^{ij}_{ab}-\tau^{ji}_{ab})^* = (\Gamma^{ab}_{ij})^*, \\
\Gamma^{jb}_{ia} &= 2\delta^j_i\gamma^b_a = \Gamma^{bj}_{ai}, \\
\Gamma^{bj}_{ia} &= -\delta^j_i \gamma^b_a = \Gamma^{jb}_{ai}.
\end{align}

\newpage
\section{Molecular geometries}
Table~\ref{geometry} lists the molecular geometries used in the main article. 

\begin{table}
\caption{Molecular geometries (Cartesian coordinates, in Bohr).}
\centering
\begin{tabular}{l l S[table-format=+1.4] S[table-format=+1.10] S[table-format=+1.10]}
\hline\hline
\ch{HF} & H & 0.0 & 0.0 & 0.0 \\
        & F & 0.0 & 0.0 & 1.7328795 \\
\hline
\ch{H2O} & O & 0.0 & 0.0            & -0.1239093563 \\
         & H & 0.0 &  1.4299372840  & 0.9832657567 \\
         & H & 0.0 & -1.4299372840  & 0.9832657567 \\
\hline
\ch{NH3} & N & 0.0    &  0.0    &  0.2010 \\
         & H & 0.0    &  1.7641 & -0.4690 \\
         & H & 1.5277 & -0.8820 & -0.4690 \\
         & H & -1.5277& -0.8820 & -0.4690 \\
\hline
\ch{CH4} & C & 0.0 & 0.0 & 0.0 \\ 
         & H & 1.2005 & 1.2005  & 1.2005 \\ 
         & H & -1.2005 & -1.2005& 1.2005 \\
         & H & -1.2005 & 1.2005 & -1.2005 \\
         & H & 1.2005 & -1.2005 & -1.2005 \\
\hline\hline
\end{tabular}
\label{geometry}
\end{table}

\newpage
\section{Ground-state energies and electric dipole moments}

Table~\ref{groundstate_properties} lists ground-state energies and the $z$-component of the electric dipole moments computed with
the d-aug-cc-pVDZ basis set for \ch{Ne} and the aug-cc-pVDZ basis set for the remaining molecules.

\begin{table}
\centering
\caption{Ground-state energies (a.u.) and electric dipole moments (a.u.)
         for \ch{Ne}, \ch{HF}, \ch{H2O}, \ch{NH3}, \ch{CH4}.}
\begin{tabular}{l l S[table-format=+3.10] S[table-format=+1.10]}
 \hline 
 \hline
  & & {$E_0$} & {$\mu_z$} \\
 \hline
 \ch{Ne}  &  CCSD & -128.7088211871 &  0.0          \\
 		  &  OMP2 & -128.7070147802 &  0.0          \\		  
		  &  CC2  & -128.7074682467 &  0.0          \\ 
 \ch{HF}  &  CCSD & -100.2615084708 & -0.7032371436 \\
 		  &  OMP2 & -100.2601905137 & -0.7005241699 \\          
		  &  CC2  & -100.2605587124 & -0.6890477615 \\
 \ch{H2O} &  CCSD & - 76.2707676433 &  0.7290920663 \\
 		  &  OMP2 & - 76.2654705768 &  0.7247294276 \\ 		  
		  &  CC2  & - 76.2655194137 &  0.7163971417 \\
 \ch{NH3} &  CCSD & - 56.4213262714 & -0.5753808611 \\
 		  &  OMP2 & - 56.4081347405 & -0.5709463975 \\
		  &  CC2  & - 56.4080164764 & -0.5688980571 \\
 \ch{CH4} &  CCSD & - 40.3941433359 &  0.0          \\
 		  &  OMP2 & - 40.3717689870 &  0.0          \\
		  &  CC2  & - 40.3716578990 &  0.0          \\
 \hline 
 \hline
\end{tabular}
\label{groundstate_properties}
\end{table}

\section{Comparison of absorption spectra from TDCC2 and LRCC2 theory}

\begin{figure}
    \centering
    \subfloat{{\includegraphics[scale=1]{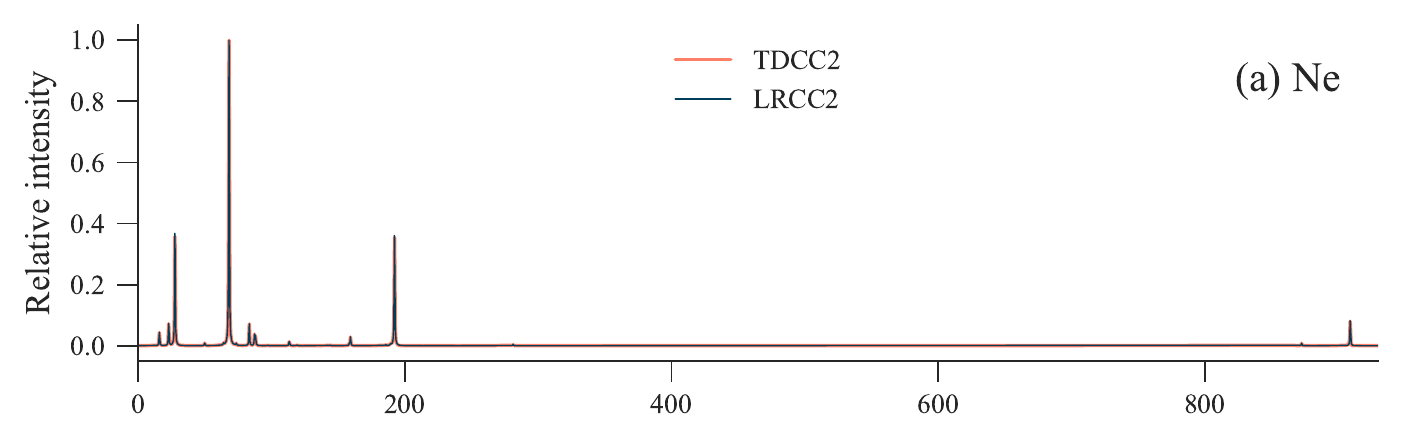} }}\\
    \vspace{-1.3\baselineskip}
    \subfloat{{\includegraphics[scale=1]{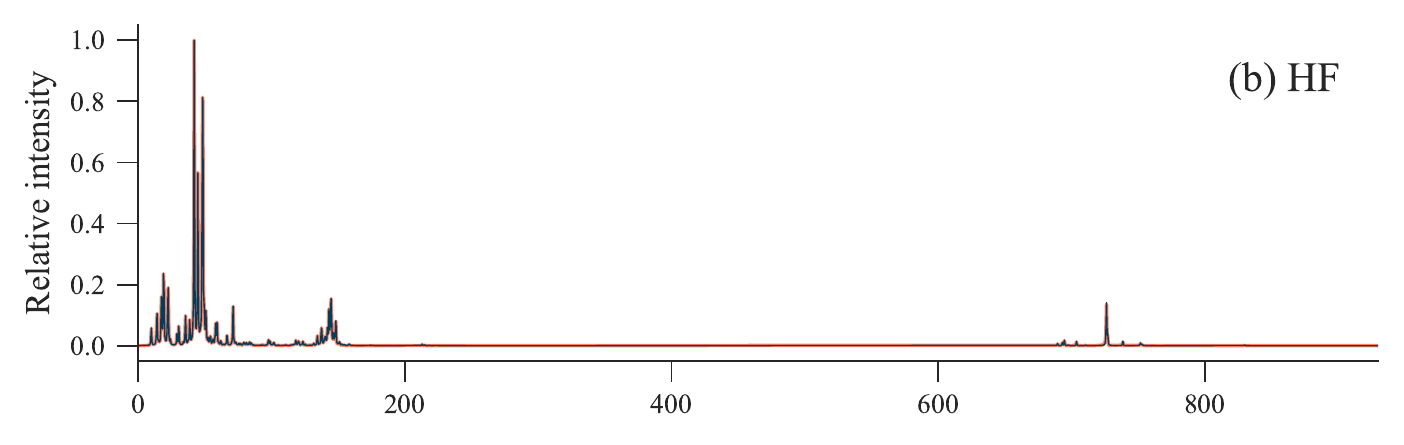} }}\\
    \vspace{-1.3\baselineskip}
    \subfloat{{\includegraphics[scale=1]{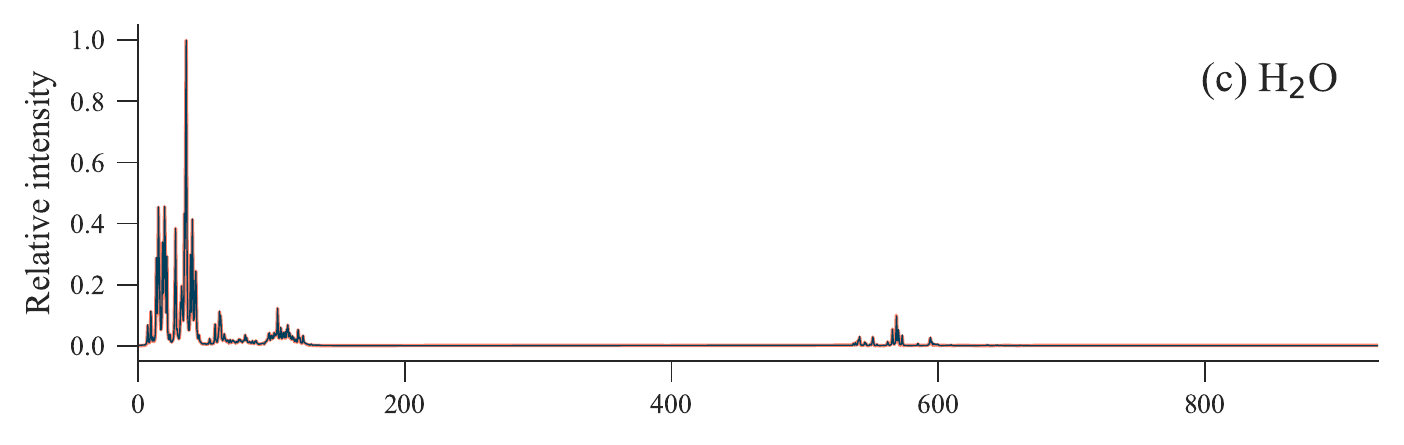} }}\\
    \vspace{-1.3\baselineskip}
    \subfloat{{\includegraphics[scale=1]{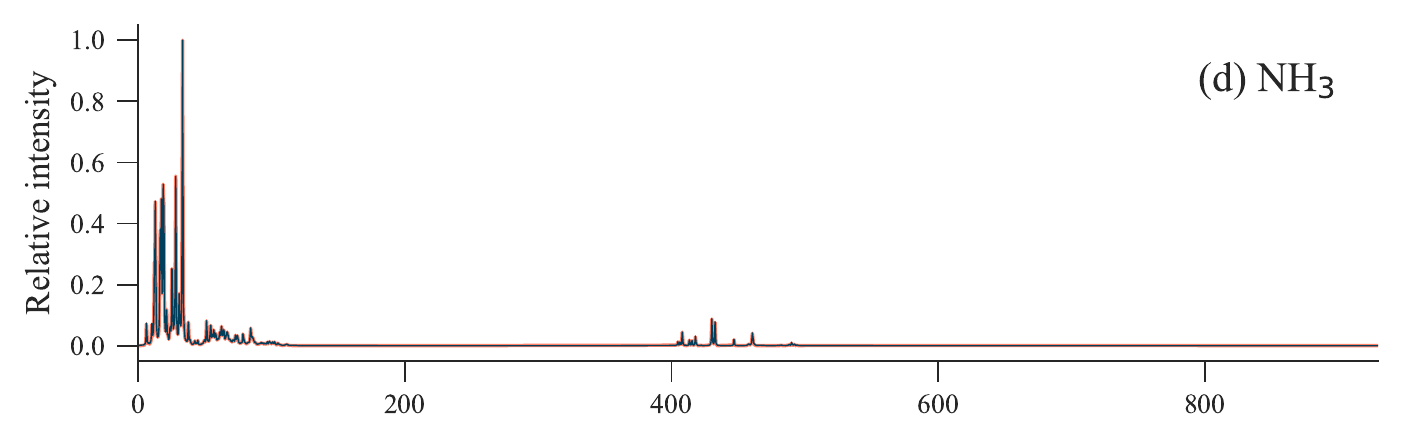} }} \\
    \vspace{-1.3\baselineskip}
    \subfloat{{\includegraphics[scale=1]{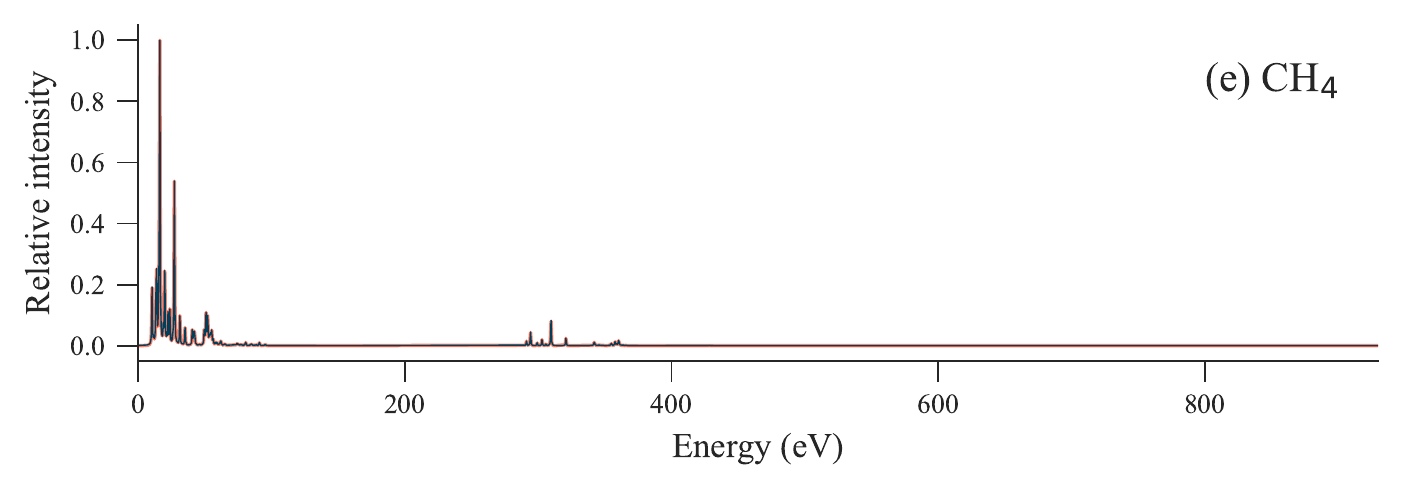} }} \\    
    \caption{Absorption spectra from TDCC2 simulations and LRCC2 calculations for \ch{Ne}, \ch{HF}, \ch{H2O}, \ch{NH3} and \ch{CH4}.}%
    \label{fig:absorption_spectra_SI}%
\end{figure}

Figure \ref{fig:absorption_spectra_SI} shows the agreement between spectra from TDCC2 simulations and
from LRCC2 calculations using the Lanczos-chain-driven algorithm described by \citeauthor{Coriani_Core_2012}~\cite{Coriani_Core_2012,Coriani2012}
as implemented in the Dalton quantum chemistry program~\cite{Aidas2014,Olsen2020}.
The same geometries and basis sets as above were used.


\bibliography{manuscript}